\documentclass{aa}

\usepackage{graphics}
\usepackage{longtable}
\usepackage{txfonts}
\usepackage{graphicx}
\usepackage{hyperref}
\usepackage{multirow}

\begin{document}

\title{First spectro-interferometric survey of Be stars}
\subtitle{I. Observations and constraints on the disk geometry and kinematics}
\authorrunning{-}
\titlerunning{-}

\author{A. Meilland\inst{1}, F. Millour\inst{1}, S. Kanaan\inst{2}, Ph. Stee\inst{1}, R. Petrov\inst{1}, K.-H. Hofmann\inst{3}, A. Natta\inst{4}, and K. Perraut\inst{5}}

   \offprints{ame@oca.eu}

\institute{
UMR 6525 CNRS H. FIZEAU UNS, OCA, Campus Valrose, F-06108 Nice cedex 2
\and
Departamento de F\'isica y Astronom\'ia, Universidad de Valpara\'iso, Chile.
\and Max-Planck-Institut f\"ur Radioastronomie, Auf dem H\"ugel 69, D-53121 Bonn, Germany
\and INAF – Osservatorio Astrofisico di Arcetri, Istituto Nazionale di Astrofisica, Largo E. Fermi 5, 50125 Firenze, Italy
\and UJF-Grenoble 1 / CNRS-INSU, IPAG UMR 5274, Grenoble, F-38041, France}
\date{Received; accepted }

   \abstract
{Classical Be stars are hot non-supergiant stars surrounded by a gaseous circumstellar disk that is responsible for the
observed infrared-excess and emission lines. The phenomena involved in the disk formation still remain highly debated.}
{To progress in the understanding of the physical process or processes responsible for the mass ejections and test the hypothesis that they depend on the stellar parameters, we initiated a survey on the circumstellar environment of the brightest Be stars.}
{To achieve this goal, we used spectro-interferometry, the only technique that  combines high spectral (R=12000) and high spatial ($\theta_{\rm min}$=4\,mas) resolutions. Observations were carried out at the Paranal observatory with the VLTI/AMBER instrument. We concentrated our observations on the Br$\gamma$ emission line to be able to study the kinematics within the circumstellar disk. Our sample is composed of eight bright classical Be stars : $\alpha$ Col, $\kappa$ CMa, $\omega$ Car, p	Car, $\delta$ Cen, $\mu$ Cen, $\alpha$ Ara, and \textit{o} Aqr.}
{We managed to determine the disk extension in the line and the nearby continuum for most targets. We also constrained the disk kinematics, showing that it is dominated by rotation with a rotation law close to the Keplerian one. Our survey also suggests that these stars are rotating at a mean velocity of V/V$_{\rm c}$\,=\,0.82\,$\pm$\,0.08. This corresponds to a rotational rate of $\Omega$/$\Omega_{\rm c}$\,=\,0.95\,$\pm$\,0.02}
{We did not detect any correlation between the stellar parameters and the structure of the circumstellar environment. Moreover, it seems that a simple model of a geometrically thin Keplerian disk can explain most of our spectrally resolved K-band data. Nevertheless, some small departures from this model have been detected for at least two objects (i.e, $\kappa$ CMa and $\alpha$ Col). Finally, our Be stars sample suggests that rotation is the main physical process driving the mass-ejection. Nevertheless, smaller effects from other mechanisms have to be taken into account to fully explain how the residual gravity is compensated.}

   \maketitle
%

\begin{table*}[!t]
\caption{\label{params} Be stars observed in this survey and their stellar parameters taken from the literature or estimated from the SED fit.}
{\centering \begin{tabular}{ccccccccccc}
\hline 
Name					&HD			& m$_{\rm K}$	&spectral 	&distance	& T$_{\rm eff}$	&v\,sin\,i			& V$_{c}$					& pol. angle		& estim. R$_\star$&estim. F$_{env}$\\
							&				&	(mag)				&	class			&(pc)			& (K)						&(km\,s$^{-1}$)	& (km\,s$^{-1}$)	& (deg)					&	(R$_\odot$)			& (in the K band)\\
\hline 
$\alpha$ Col	&37795	&2.8					&B7IV						&80$\pm $2&12963$\pm$203	&192$\pm$12			&355$\pm$23				&109				&5.8							&0.25\\
$\kappa$ CMa	&50013	&3.5					&B1.5V					&202$\pm$5&24627$\pm$590	&244$\pm$17			&535$\pm$39				&106				&5.9							&0.47\\
$\omega$ Car	&89080	&3.5					&B8III					&105$\pm$1&11720$\pm$431	&245$\pm$13			&320$\pm$17				&38					&6.2							&0.20\\
p	Car					&91465	&3.2					&B4V						&148$\pm$9&17389$\pm$415	&285$\pm$20			&401$\pm$28				&68					&6.0							&0.45\\
$\delta$ Cen	&105435	&2.7					&B2IV						&127$\pm$8&22360$\pm$589	&263$\pm$14			&527$\pm$29				&137				&6.5							&0.45\\
$\mu$ Cen 		&120324	&4.0					&B2IV-V					&155$\pm$4&22554$\pm$661	&166$\pm$10			&508$\pm$32				&-					&5.5							&0.37\\
$\alpha$ Ara	&158427	&2.5					&B3IV						&82$\pm$5	&18044$\pm$310	&305$\pm$15			&477$\pm$24				&166				&5.5							&0.56\\
\textit{o} Aqr&209409	&4.6					&B7IV						&133$\pm$4&12942$\pm$402	&282$\pm$20			&391$\pm$27				&6					&4.0							&0.31\\
\hline 
\end{tabular}\par}
\end{table*}

\section{Introduction} 

Classical Be stars are hot non-supergiant stars that have at least once exhibited  the so called ``Be-phenomenon", i.e. Balmer lines in emission and IR-excess originating from a dense gaseous circumstellar environment. A generally accepted scheme is the presence of two distinct regions in the envelope: a dense equatorial disk dominated by rotation that is responsible for most of the line emission and IR-excess, and a more diluted radiatively driven polar wind  (Lamers \& Waters 1987) with terminal velocities on the order of several hundreds of km.s$^{-1}$ (Marlborough 1982).

However, the physical process or processes responsible for the mass ejection and reorganization of matter in the circumstellar 
environment are still highly debated. The relative effect of rotation, radiative pressure, pulsation, and binarity still need to be quantified. For instance, classical Be stars are known to be fast-rotators, but estimates of their rotational velocities range from 50 to 100$\%$ of the critical velocity which means that rotation alone might not explain the mass-ejection for all cases. Cranmer (2005) showed in a statistical study of the velocity of 462 classical Be stars that the hottest Be stars (i.e., T$_{\rm eff}>$18000K) have a large spread of rotational velocities, whereas the cooler Be stars are more likely to be critical rotators. Moreover, Abbott (1979) showed that the radiative pressure alone can initiate mass ejection only for the earliest Be stars. Consequently, it is not yet clear if Be stars can be considered as a homogeneous group of stars in terms of mass-ejection processes (Stee \& Meilland 2009).

An efficient way to test these hypotheses is to constrain the geometry and kinematics of the Be stars circumstellar environments because they mainly depend on the mechanisms responsible for the mass-ejection. This can only be performed with long-baseline interferometry with sufficient spectral resolution, as shown by Meilland et al. (2007a ; 2007b; 2011) using the VLTI/AMBER on the stars $\alpha$ Ara, $\kappa$ CMa, and $\delta$ Sco, or Delaa et al. (2010) using the CHARA/VEGA instrument on 48 Per and $\Phi$ Per. In a few cases, the stellar photosphere can also be resolved by the interferometer, allowing measuring the photosphere's flattening. In this case the physical stellar parameters can be determined, as for Achernar (Domiciano de Souza et al. 2003). For this specific object, a clear signature of a polar wind was also detected using the same VLTI/VINCI dataset by Kervella et al. (2006).

Before the availability of a new generation of instruments coupling high spectral and spatial resolutions, interferometric studies were conducted in  photometric bands to measure extension and flattening of the circumtellar environments of many Be stars. For instance, $\gamma$ Cas and $\psi$ Per were observed by \cite{Tycner06} with the NPOI interferometer in the H$\alpha$ domain. They found that a uniform disk or a ring-like model were inconsistent with their data, whereas a Gaussian model well fitted the measurements. $\gamma$ Cas' disk was also consistent with the orbital parameters already published. Nevertheless, higher precision binary solutions were mandatory to test for a possible disk truncation by the secondary. The disk of $\psi$ Per was found to be truncated by a companion as already predicted by \cite{Waters86}.

$\gamma$ Cas, $\psi$ Per, $\zeta$ Tau and $\kappa$ Dra were also observed with the CHARA interferometer in the K-band by \cite{Gies07}. Using Gaussian elliptical fits of visibilities, they found that the disk size in the K-band was smaller than in H$\alpha$ owing to a higher H$\alpha$ opacity and relatively higher neutral hydrogen fraction with increasing disk radius. All these Be stars are known binaries, and this binarity effect was found to be more significant for $\psi$ Per and $\kappa$ Dra. 
 
\cite{Tycner08} observed $\chi$ Oph with the NPOI interferometer and obtained a good fit of the H$\alpha$ emitting disk with a circularly symmetric Gaussian, favoring the hypothesis that this object is seen under a very low inclination angle.

Using the mid-infrared VLTI/MIDI instrument, \cite{Meilland09} have observed seven classical Be stars between 8 and 12$\mu$m. They found that the size of the disk does not vary strongly with wavelength within this spectral domain, which is a very different conclusion compared to the results for B[e] stars where increasing  sizes as a function of wavelength where found (Millour et al 2009; Meilland et al. 2010; Borges Fernandes et al. 2011). Moreover, the size of $\alpha$ Arae's disk was found to be identical at 2, 8, and 12 $\mu$m, which might be because of to a disk truncation by a companion. Finally, it seems from Meilland et al.'s (2009) studies that the disks of late type Be stars might be smaller than those of the early type.

Long-baseline interferometry is also a powerful technique for detecting companions, which was shown in Meilland et al. (2008) for the Be star $\delta$ Cen and Millour et al. (2009) for the B[e] star HD\,87643. Moreover, in a theoretical study of the formation and dissipation of the Be stars' equatorial disks, Meilland et al. (2006) showed that interferometric follow-up of these events is the best-suited technique to deduce the physical parameters of the system. However, Kanaan et al. (2008) showed that, for Achernar, coupling spectroscopic follow-up and large-band interferometric observations at one epoch was sufficient to roughly understand the geometry and kinematics of this star.

To progress in the understanding of Be stars, we initiated an observational campaign of the brightest, closest objects using the VLTI/AMBER (Petrov et al. 2007; Robbe-Dubois et al. 2007) and VLTI/MIDI (Leinert et al. 2003) instruments for the southern stars and the CHARA/VEGA  (Mourard et al. 2009) for the northern ones.  In this paper, we present new VLTI/AMBER spectro-interferometric observations of eight classical Be stars :  $\alpha$ Col, $\kappa$ CMa, $\omega$ Car, p	Car, $\delta$ Cen, $\mu$ Cen, $\alpha$ Ara,  and \textit{o} Aqr.

The paper is organized as follows. In Sect. 2 we briefly introduce each target and constrain their physical parameters from various sources in the literature. The observations and data reduction process are then presented in Sect. 3, and in Sect. 4 we provide a qualitative analysis of the reduced data for each object. Then, in Sect. 5, the data are analyzed with various ``toy models''. Finally, these results are discussed in Sect. 6 and a short conclusion is drawn in Sect. 7.

\section{Our Be star sample}

Considering the actual limiting magnitude of the VLTI/AMBER instrument, i.e. H=K=5 for an unresolved source observed with the 1.8m auxiliary telescopes in medium or high spectral resolution modes, about 30 classical Be stars are observable. However, to constrain the circumstellar disk kinematics efficiently, we decided to exclude targets with weak emission lines or transient disks. Moreover, for this first survey we mainly concentrated on targets with m$_K\leq$4. Finally, we also decided to include our first attempt to observed a fainter target, i.e. \textit{o} Aqr (m$_K$=4.6).

Because one of our goals is to determine whether or not the Be phenomenon depends on the basic stellar parameters, we tried to select targets with the widest range of spectral classes possible. Finally, the stars selected in our sample, i.e. $\alpha$ Col, $\kappa$ CMa, $\omega$ Car, p	Car, $\delta$ Cen, $\mu$ Cen, $\alpha$ Ara,  and \textit{o} Aqr, have spectral types ranging from B1.5 to B8, and luminosity class of III, IV and V. Table~\ref{params} briefly describes the targets:

\begin{itemize}
\item Spectral class and the K-band magnitude (m$_{\rm K}$) are taken from the CDS databases. 
\item Distance (d) is derived from van Leeuwen (2007).
\item Effective temperature (T$_{\rm eff}$), v\,sin\,i, and critical velocity (V$_{\rm c}$) are taken from Fr\'emat et al. (2005)
\item Polarization measurement is taken from Yudin (2001)
\item	Stellar radius (R$_\star$) and K-band environment relative flux (F$_{\rm env}$) are estimated by fitting the spectral energy distribution (SED) using reddened Kurucz (1979) models for stellar atmospheres and using stellar parameters ($T_{\rm eff}$ and $g_{eff}$) from Fr\'emat et al. (2005). The SED is first reconstructed using photometric and spectro-photometric measurements from the ultraviolet (IUE specta) to the far-infrared (IRAS data). To avoid contamination from the circumstellar flux, the fit of the stellar contribution to the flux is made from the ultraviolet to the visible.(see Meilland et al. 2009 for more details). 
\end{itemize}

\begin{table*}[!t]
\caption{\label{log}VLTI/AMBER observing log. }
{\centering \begin{tabular}{ccccccccc}
\hline 
Obs. time&Telescopes&Length&Position angle &Instrument&DIT &Coherence&Seeing &Calibrators \\
(UTC)& conf.&(m) &($^o$) &mode&(s)&(ms)&(")&(HD)\\
\hline 
\multicolumn{7}{c}{\large{$\alpha$ Col}}\\
\hline 
2008-01-06 03:39&K0-G1-A0& 88.6/ 90.5/126.2&-150.1/ -59.7/-104.3&LR-K-F&0.05&5.4&1.10&81188\\
2010-01-09 00:48&D0-H0-K0& 60.2/ 30.1/ 90.3&  51.4/  51.4/  51.4&HR-K-F&5.00&3.1&1.25&34642\\
2010-01-12 00:48&G1-D0-H0& 27.3/ 50.0/ 66.6& -68.0/  -3.8/ -25.4&HR-K-F&5.00&3.9&1.06&-\\
2010-01-20 01:35&K0-G1-A0& 90.0/ 88.7/127.5&-157.9/ -68.9/-113.8&HR-K-F&5.00&5.3&0.85&34642\\
2010-01-20 01:56&K0-G1-A0& 89.9/ 89.6/128.0&-156.1/ -67.1/-111.7&HR-K-F&5.00&3.8&1.30&34642\\
\hline 
\multicolumn{7}{c}{\large{$\kappa$ CMa}}\\
\hline
2008-12-18 03:34&U1-U3-U4&102.2/ 55.6/125.3&  17.1/  96.1/  42.9&HR-K-F&1.00&5.3&0.63&-\\
2008-12-18 07:32&U1-U3-U4& 94.5/ 58.9/115.3&  43.9/ 129.2/  74.4&HR-K-F&1.00&4.9&0.64&-\\
2008-12-20 03:37&K0-G1-A0& 90.5/ 81.4/121.3&-167.9/ -77.5/-125.7&LR-K-F&0.25&4.4&0.78&40805\\
2008-12-24 03:50&K0-G1-A0& 90.5/ 84.8/124.2&-164.3/ -74.5/-121.2&LR-K-F&0.05&4.6&0.79&27442, 57299\\
2008-12-24 05:07&K0-G1-A0& 89.9/ 90.1/128.0&-154.7/ -65.3/-109.9&LR-K-F&0.05&3.5&1.00&48305, 57299\\
2008-12-24 08:01&K0-G1-A0& 80.9/ 82.8/102.2&-139.8/ -37.0/ -87.6&LR-K-F&0.05&5.4&0.66&48305, 57299\\
2010-01-09 01:39&D0-H0-K0& 57.8/ 28.9/ 86.6&  46.2/  46.2/  46.2&HR-K-F&5.00&3.1&1.26&34642\\
2010-01-18 00:24&H0-G0-E0& 27.4/ 13.7/ 41.0&-142.8/-142.8/-142.8&HR-K-F&5.00&3.6&0.71&54173\\
\hline 
\multicolumn{7}{c}{\large{$\omega$ Car}}\\
\hline

2008-12-21 04:49&K0-G1-A0& 74.2/ 72.9/128.0& 175.8/-125.0/-154.9&HR-K-F&5.00&3.8&0.78&75063, 69596\\
2008-12-21 07:37&K0-G1-A0& 72.1/ 83.0/126.6&-156.4/ -85.7/-118.2&HR-K-F&5.00&3.4&0.89&69596\\
2008-12-24 06:33&K0-G1-A0& 73.7/ 79.3/127.7&-167.4/-100.5/-132.6&LR-K-F&0.05&3.0&1.20&98134, 57299\\
\hline 
\multicolumn{7}{c}{\large{p Car}}\\
\hline
2008-12-23 07:31&K0-G1-A0& 79.3/ 82.3/127.8&-161.2/ -85.7/-122.7&MR-K-F&1.00&3.3&1.11&94286, 69596\\
2009-03-22 03:59&K0-G1-A0& 71.7/ 89.6/120.8&-138.0/ -54.4/ -90.6&MR-K-F&1.00&6.0&0.70&94286\\
\hline 
\multicolumn{7}{c}{\large{$\delta$ Cen}}\\
\hline
2009-03-21 04:18&K0-G1-A0& 83.9/ 87.9/127.4&-153.8/ -69.5/-110.4&MR-K-F&1.00&10.9&0.74&110458\\
2010-01-20 08:12&K0-G1-A0& 84.1/ 87.7/127.5&-154.5/ -70.4/-111.4&LR-HK&0.05&2.6&1.27&110458\\
2011-05-19 02:41&U1-U2-U4& 46.4/ 82.3/111.4&  43.8/ 106.6/  84.8&LR-HK&0.025&1.3&1.47&103513\\
\hline 
\multicolumn{7}{c}{\large{$\mu$ Cen}}\\
\hline
2011-06-26 23:21&K0-A1-G1&128.9/ 78.1/ 88.2&-121.5/ 100.6/-157.9&HR-K-F&6.00&1.6&0.90&128488\\
2011-06-30 00:19&D0-I1-G1& 82.3/ 44.5/ 70.8& 104.8/-134.5/ 137.5&HR-K-F&5.00&3.4&0.94&128488\\
\hline 
\multicolumn{7}{c}{\large{$\alpha$ Ara}}\\
\hline
2007-07-28 05:51&G1-D0-H0& 71.5/ 44.7/ 55.4&  -5.4/ 123.9/  33.3&LR-K&0.025&3.5&0.44&177716, 164371\\
2007-04-13 05:47&H0-G0-E0& 30.8/ 15.4/ 46.3&-143.4/-143.4/-143.4&LR-K&0.025&3.0&0.63&124454\\
2007-06-09 07:11&K0-G1-A0& 74.2/ 90.0/112.3&-135.8/ -41.6/ -82.8&LR-K&0.025&2.7&0.57&166460, 164371\\
2007-06-09 08:02&K0-G1-A0& 68.3/ 88.9/102.5&-130.4/ -30.7/ -71.7&LR-K&0.025&1.6&0.95&166460, 164371\\
2007-06-06 07:30&K0-G1-A0& 72.5/ 89.6/109.5&-134.2/ -38.5/ -79.7&LR-K&0.025&1.3&1.38&124454, 164371\\
2007-06-06 08:50&K0-G1-A0& 63.4/ 88.0/ 95.0&-126.9/ -22.7/ -63.0&LR-K&0.025&1.3&1.42&164371\\
2007-04-14 06:22&H0-G0-E0& 31.4/ 15.7/ 47.2&-133.5/-133.5/-133.5&LR-K&0.025&2.9&0.67&164371\\
2007-05-16 07:34&H0-D0-A0& 61.1/ 30.6/ 91.7& -96.4/ -96.4/ -96.4&LR-K&0.025&3.0&1.25&124454, 164371\\
2007-06-10 07:44&K0-G1-A0& 70.2/ 89.2/105.7&-132.0/ -34.1/ -75.2&LR-K&0.025&2.3&0.58&164371, 21201\\
2007-05-17 02:34&H0-D0-A0& 60.2/ 30.1/ 90.3&-156.3/-156.3/-156.3&LR-K&0.025&3.0&0.79&124454\\
2007-05-17 09:53&H0-D0-A0& 49.3/ 24.6/ 73.9& -67.2/ -67.2/ -67.2&LR-K&0.025&1.9&1.53&164371\\
2011-06-30 02:04&D0-I1-G1& 78.2/ 45.8/ 62.7&  82.6/-150.7/ 118.4&HR-K-F&5.00&3.7&0.85&163145, 152786\\
2011-06-30 03:03&D0-I1-H0& 81.4/ 37.2/ 63.8&  93.9/ -36.5/  67.6&HR-K-F&5.00&2.9&1.08&163145, 152786\\
2011-06-30 03:37&D0-I1-H0& 82.2/ 38.1/ 63.2& 100.4/ -31.9/  73.9&HR-K-F&5.00&2.9&1.05&163145, 152786\\
2011-06-30 04:28&D0-I1-H0& 82.0/ 39.0/ 61.2& 110.1/ -24.7/  83.3&HR-K-F&5.00&2.4&1.27&163145, 152786\\
2011-06-30 05:03&D0-I1-H0& 81.0/ 39.4/ 58.9& 117.5/ -19.2/  90.2&HR-K-F&5.00&2.9&1.03&163145, 152786\\
2011-06-30 06:00&D0-I1-H0& 78.5/ 39.8/ 54.4& 129.4/ -10.7/ 101.3&HR-K-F&5.00&2.9&1.02&163145, 152786\\
2011-06-30 06:38&D0-I1-H0& 76.4/ 39.9/ 50.6& 138.1/  -4.7/ 109.7&HR-K-F&5.00&3.2&0.92&163145, 152786\\
\hline 
\multicolumn{7}{c}{\large{\textit{o} Aqr}}\\
\hline
2011-06-30 10:08&D0-I1-H0& 67.4/ 34.3/ 58.8& 102.8/ -16.5/  72.2&HR-K-F&5.00&2.9&1.02&209926\\
\hline 
\end{tabular}\par}
\end{table*}

\begin{figure*}[!tbh]
\centering   
\vspace{-0.2cm}
\includegraphics[width=0.24\textwidth]{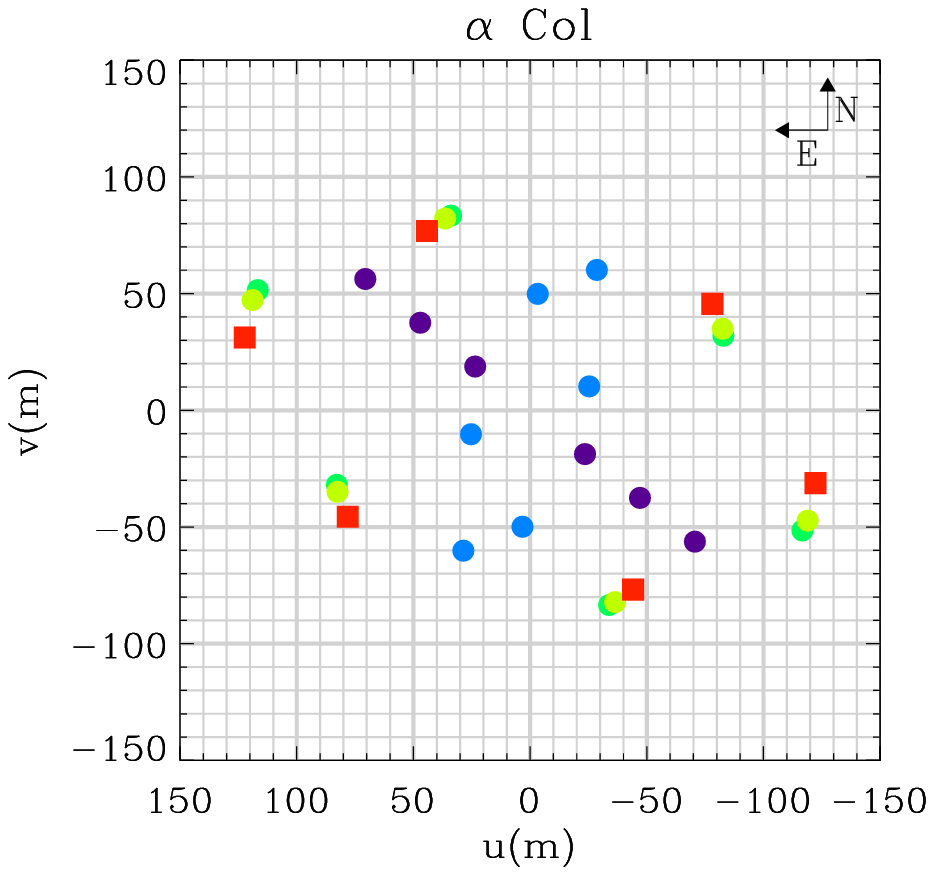}
\includegraphics[width=0.24\textwidth]{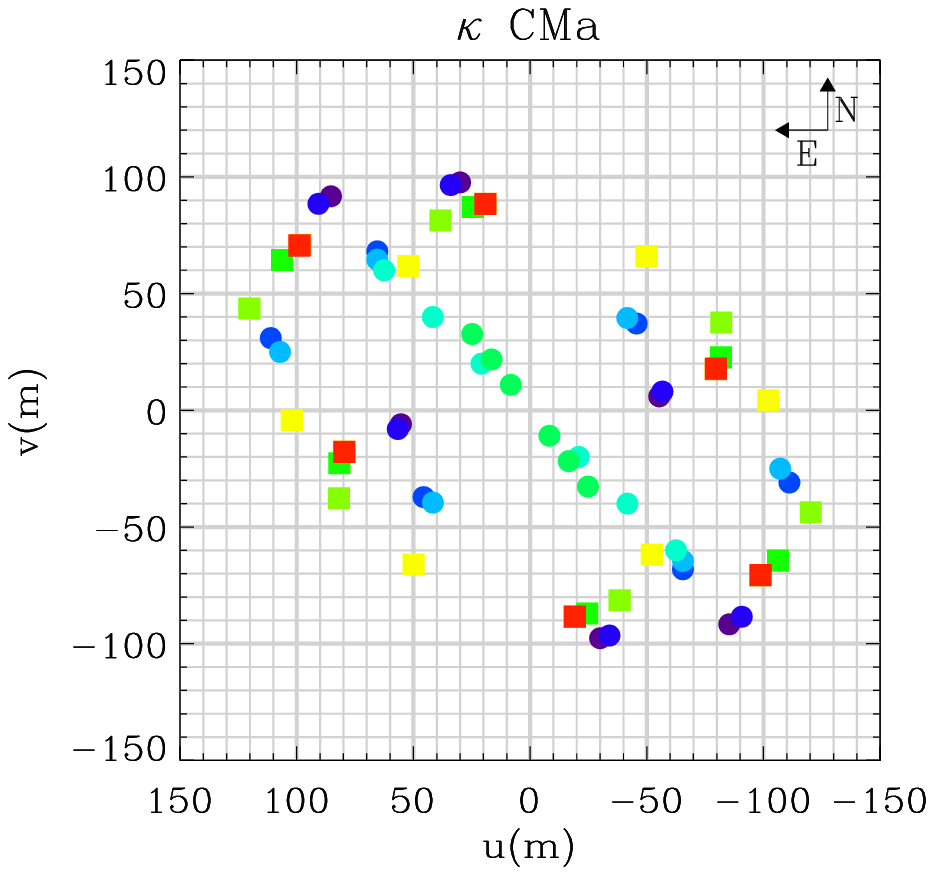}
\includegraphics[width=0.24\textwidth]{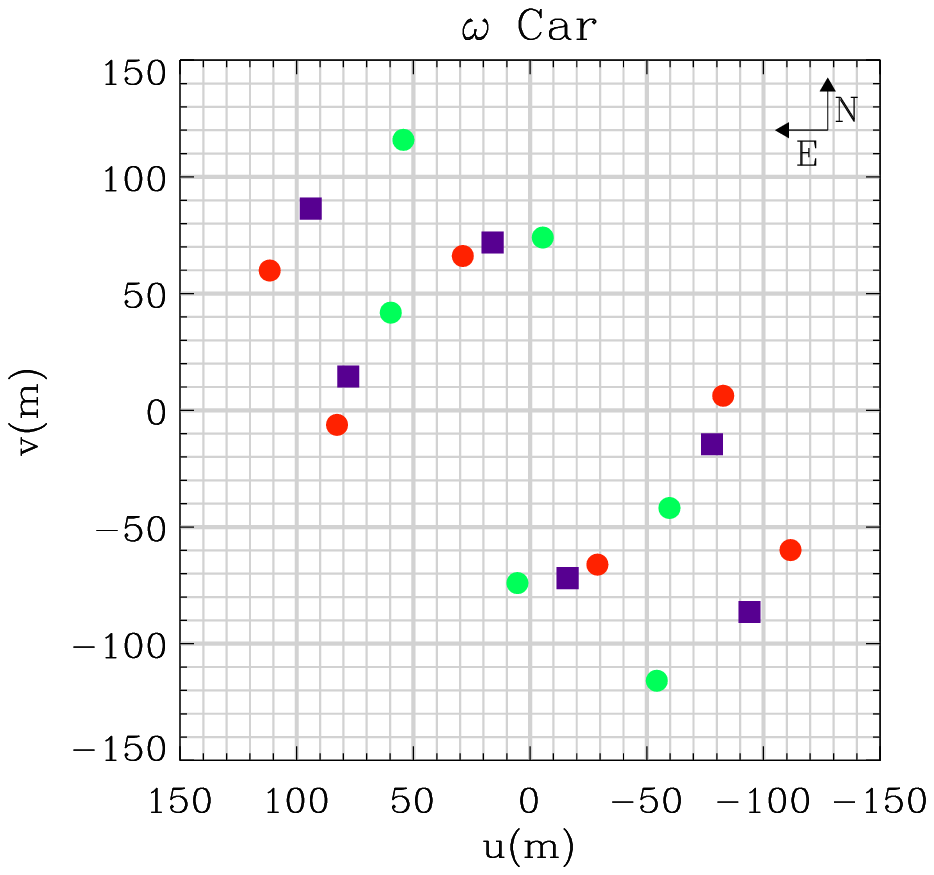}
\includegraphics[width=0.24\textwidth]{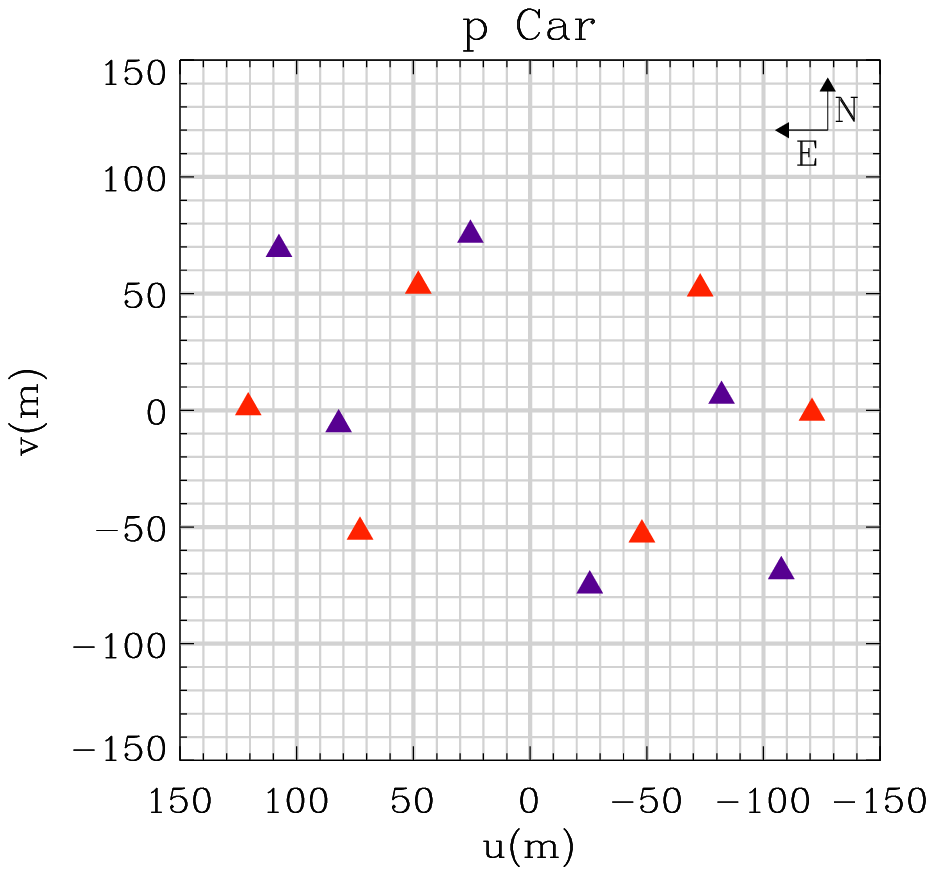}\vspace{-0.5cm}
\includegraphics[width=0.24\textwidth]{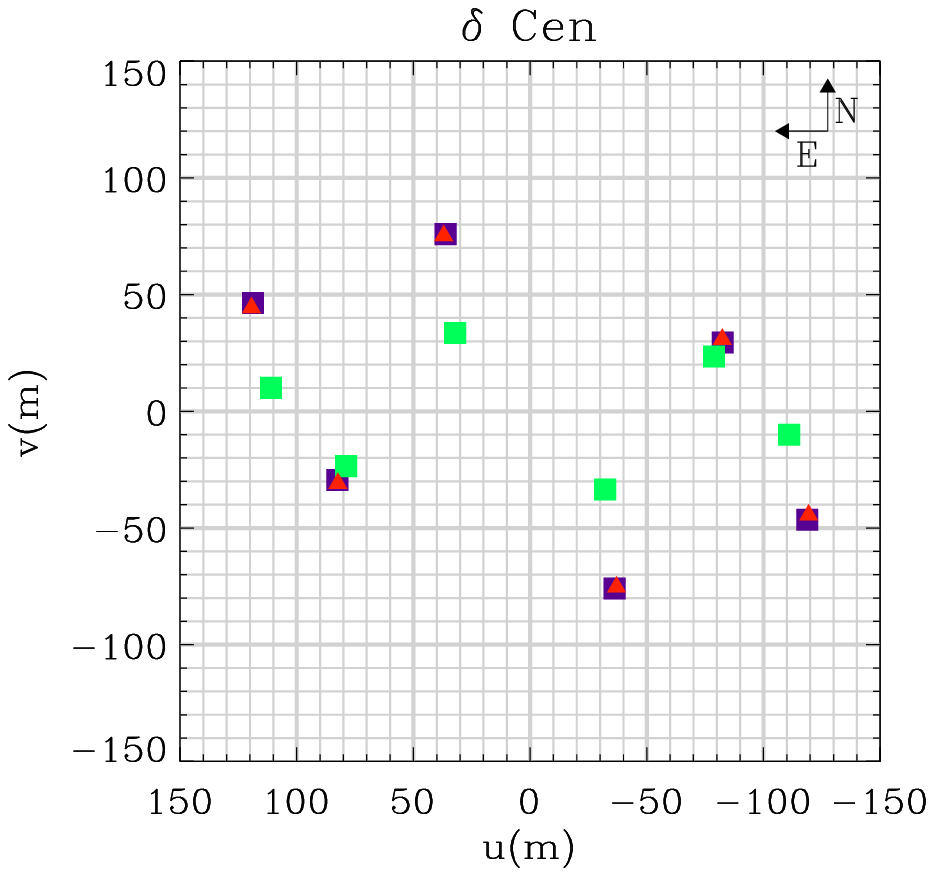}
\includegraphics[width=0.24\textwidth]{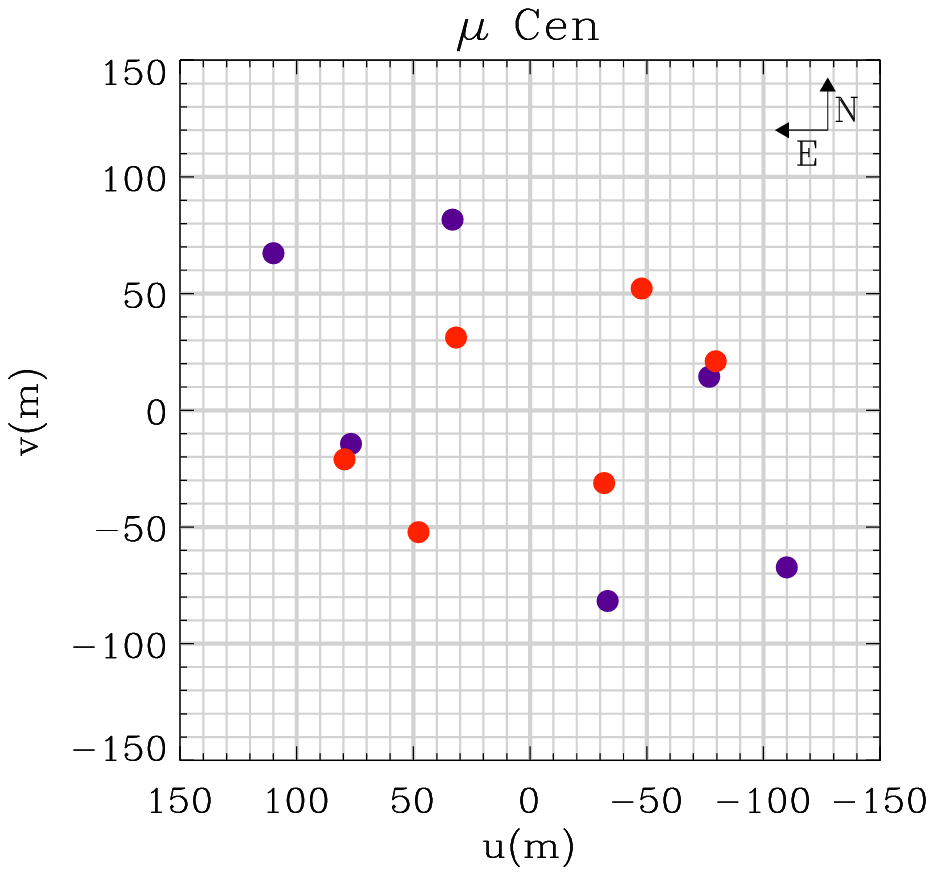}
\includegraphics[width=0.24\textwidth]{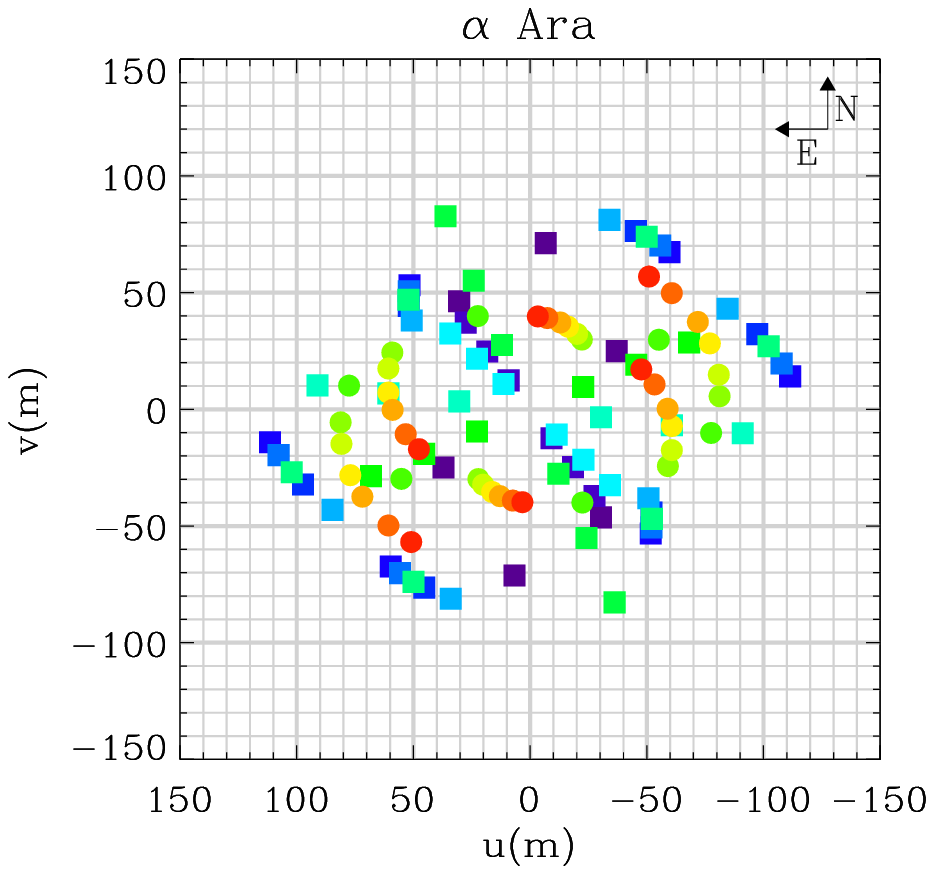}
\includegraphics[width=0.24\textwidth]{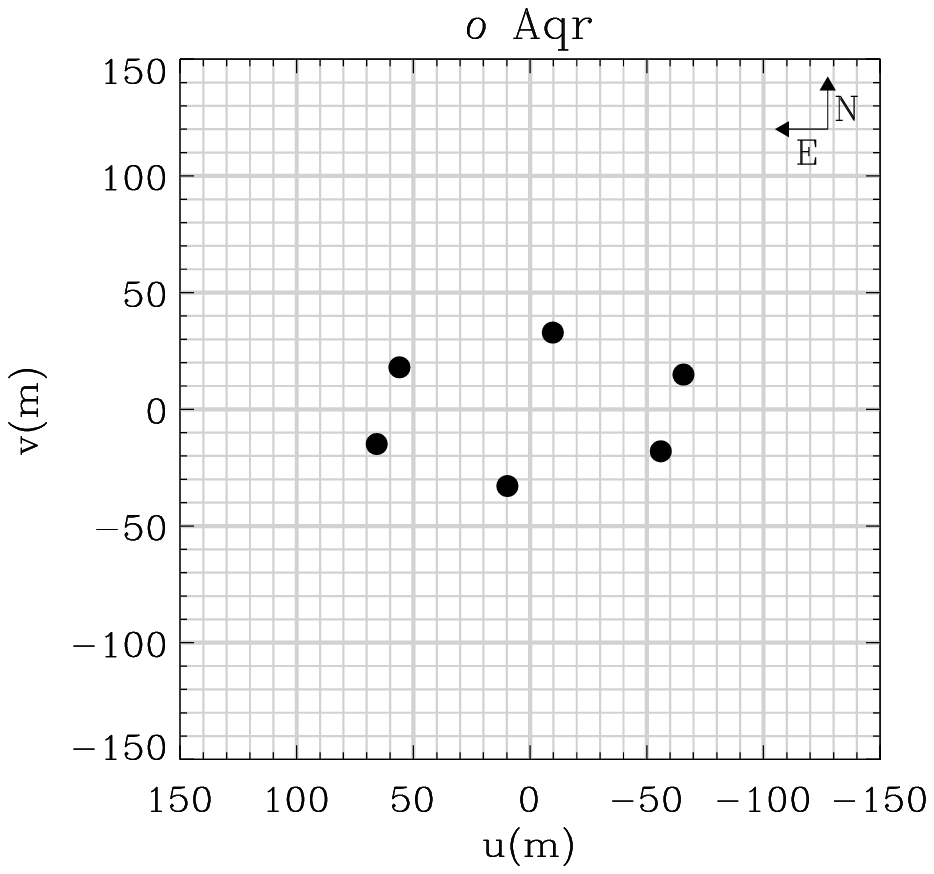}
\caption{(u,v) plan coverage obtained for the observed Be stars. LR mode observations are plotted as squares, MR ones as triangles, and HR ones as circles. Each observation, i.e. three baseline measurements, is plotted with a different color.}
\label{uv}
\end{figure*}

\section{Observations and data reduction process}

The VLTI/AMBER  observations of classical Be stars were carried out at Paranal Observatory between 2007 and 2011. Most observations were executed using the 1.8 meters movable auxiliary telescopes (AT), a few others were made using 8.0m  unit telescopes (UT). The VLTI/AMBER observing log for the eight targets and their corresponding interferometric calibrators are presented in Table~\ref{log}. The (u,v) plan coverage for each target is plotted in Fig~\ref{uv}.

Because the observations were spread over a long period of time, starting soon after the opening of AMBER to the scientific community, the data are quite inhomogeneous, both in terms of quality and observing modes. AMBER offers three different spectral dispersion modes: the low-resolution (LR) with R=$\lambda$/$\delta\lambda$$\simeq$35, the medium-resolution (MR) with R$\simeq$1500, and the high-resolution (HR) with R$\simeq$12000 . All these modes were used during our observing campaign, depending on their availability at the epoch of observations, on the brightness of the target, and on the seeing conditions. The MR and HR observations were centered on the Br$\gamma$ emission line to enable the study of the circumstellar gas kinematics through the Doppler effect.

Most of the data have benefited from the installation of the fringe tracker FINITO that enables longer integration time by stabilizing the fringes. The detector integration time (DIT) ranges from 25\,ms for bright targets in LR mode without FINITO to several seconds for observations in HR mode with FINITO. Under good seeing conditions (i.e. seeing $<$0.8'') FINITO allows one to obtain a significantly higher signal-to-noise ratio (S/N) both for short and long integration time, significantly enhancing the quality of the data for all observing modes.

Data were reduced using the VLTI/AMBER data reduction software, i.e., \texttt{amdlib v3.0.3b1} (see Tatulli et al. 2007 and Chelli et al. 2009 for detailed informations on the AMBER data reduction). We selected individual exposures with the standard selection criteria (Millour et al. 2007). We rejected  80$\%$ of frames with the lowest S/N. For observations in LR mode we also rejected the frames with a piston larger than 10$\mu$m as well as frames with a flux ratio between the beams higher than 3.

The interferometric observables (visibility, differential phase, and closure phase) were then averaged and calibrated. For this last step we used scripts described in Millour et al. (2007) that are now part of the standard amdlib package. The calibration process includes an estimation of the calibrators' size and their uncertainties from various catalogs, a determination of their transfer function and their evolution during the whole night, and a computation of the calibrated visibilities and phases. The final errors on the measurements include uncertainties on the calibrators' diameter, the atmosphere transfer function fluctuations and intrinsic errors on the measurements.

\section{A qualitivative analysis of our dataset}

The Br$\gamma$ line for all the observed targets is clearly in emission (See Fig~\ref{lines} and Table~\ref{dbpeak} for a summary of their main spectral characteristics). In almost all cases, the MR and HR data also exhibit a drop of visibility in the emission line caused by a variation of the circumstellar environment extension and relative flux between the continuum and the line, as already explained in Meilland et al. (2007a). They also exhibit an ``S'' shape or more complex variations of the differential phase in the line, and the HR data clearly show that some visibility variations are ``W'' shaped (See e.g. Fig. 3). These characteristics are clear evidence of a rotating equatorial disk as described in Meilland et al. (2011) for $\delta$~Scorpii. The MR and HR data for all targets are presented in Figs.~\ref{alpcol-vis} to \ref{omiaqr-vis}.

We note that for the HR observations executed before the replacement of a disturbing optical element in front of the VLTI \textit{InfraRed Image Sensor} (IRIS) in 2010, some instrumental modulations with a very high frequency are seen in the HR data. This is especially the case for $\omega$ Car. To enhance the data quality, we decided to filter these modulations with a standard Fourier-transform low-pass-filter technique.

Some targets are also partly resolved in the K-band continuum. However, considering the uncertainties on the calibrated absolute visibilities, it is difficult to determine an accurate extension of the circumstellar disk in the continuum for all targets. Therefore, we decided for efficiency to use only the differential visibility (i.e. visibility of each spectral channel divided by the mean visibility) for our kinematics study and the absolute continuum visibility to determine the K-band extension when possible.

Among our sample we did not detect any new companion, and $\delta$ Cen remains the only star for which a companion was visible in the interferometric signal, as already described by Meilland et al. (2008). The individual comments for the observed objects are the following:
\begin{figure}[!t]
\centering   
\includegraphics[width=0.43\textwidth]{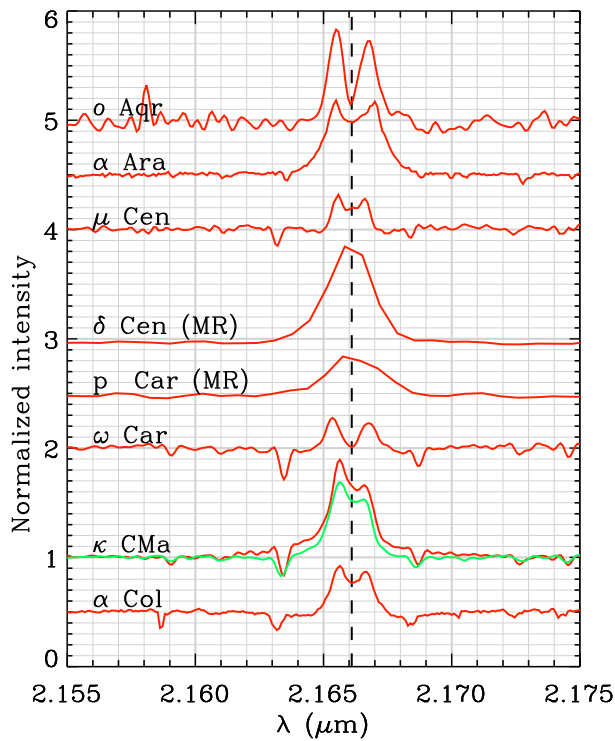}
\caption{Br$\gamma$ spectra from our VLTI/AMBER observations. The data were taken in HR mode for all observations expect for p Car and $\delta$ Cen (MR mode). For $\kappa$ CMa the red and green lines represent the 2008 and 2010 observations, respectively. }
\label{lines}
\end{figure}

\begin{itemize}
\item \noindent\underline{$\alpha$ Col} : We obtained three measurements in HR and one in LR. The data S/N is quite high and the uncertainties on the differential quantities are on the order of a few percents. The target is clearly resolved in the line and the measurements in HR clearly exhibit the typical visibility and phase variations of a rotating disk with the major axis roughly perpendicular to the polarization measurement obtained by Yudin 2001 (see Table~\ref{params}). The quasi-symmetric double-peaked line profile indicates that the object is seen from an intermediate inclination angle and that no major inhomogeneity is present in the disk. The object is also partly resolved in the continuum, with a visibility of about 0.9$\pm$0.05 for the longest baselines.

\item\underline{$\kappa$ CMa} : This star was observed for the first time in December 2008 and we have obtained two measurements in HR and five in LR. It was observed again in January 2010 and two new measurements in HR were acquired. The S/N is quite high except for the last observations in 2010. All HR measurements at the two epochs show the typical visibility and phase variations of a rotating disk. The disk major axis seems to be roughly perpendicular to the polarization measurement. The double-peaked line profile and some visibility and phase variations are strongly asymmetric. These asymmetries probably stem from an inhomogeneity in the disk that was already described in Meilland et al. (2007b). We note that the line profile violet-over-red (V/R) peaks ratio did not significantly vary between our 2008 and 2010 observations. The LR data show that the star is at least partly resolved in the continuum, but the calibration is not accurate enough to fully constrain the continuum extension.

\item\underline{$\omega$ Car} : We obtained two measurements in HR and one measurement in LR. Although the S/N is lower compared to $\alpha$~Col, a visibility and phase signals typical of a rotating disk are also present in the data. The major axis does not seem to be aligned or perpendicular to the polarization measurement. Because the line profile is double-peaked with a narrow shell line at its center, the star is probably seen from a high inclination angle. The LR data clearly indicate that the object is partly resolved in the continuum, i.e. V$\sim$0.85 for the longest baseline. We note that the second HR measurement cannot be calibrated accurately.

\item\underline{p Car} : We obtained two measurements in MR, both showing a bright emission line, a drop of visibility, and ``S" shaped phase variations in the line. Considering the phase amplitude and the lack of spectral resolution, it is hard to qualitatively determine the position of the major axis of the equatorial disk. We note that the profile and phase variations are asymmetric. This is probably cause by an inhomogeneity such as a one-armed oscillation, as proposed by Okazaki (1997). The object is barely resolved in the continuum (V$\sim$0.9 for the longest baselines). 

\item\underline{$\delta$ Cen} : Previous VLTI/AMBER LR and MR observations published by Meilland et al. (2008) showed the binarity of this object. The authors found a contribution of the companion to the total K-band flux of 7$\%$ and a separation of 68.7\,mas. To check their results and constrain the system orbit, the star was observed again in  2009, 2010, and 2011. In the 2009 MR data a bright Br$\gamma$ emission line, a drop of visibility and ``S" shaped phase variations are clearly visible. These data as well as the 2010 LR ones also contain modulations cause by the binarity of the object. We note that no obvious modulation is seen in the 2011 LR dataset. This may mean that the separation is quite small, on the order of a few milli-arcseconds.

\begin{table}[!t]
\caption{Br$\gamma$ line characteristics\label{dbpeak}}
\centering \begin{tabular}{ccccc}
\hline
Star							&EW				&Peaks separation			&V/R			&Remarks\\
									&$\AA$		&$\AA$ (km\,s$^{-1}$)	&					&				\\
\hline\hline
$\alpha$ Col			&6.1			&9.9 (138)						&1.04		&-				\\
$\kappa$ CMa(2008)&16.8			&9.9 (138)						&1.13		&-				\\
$\kappa$ CMa(2010)&11.6			&10.4 (144)						&1.10		&-				\\
$\omega$ Car			&3.5			&14.2 (196)						&1.04		&Be-Shell\\
p Car							&8.2			&$\sim$15 (200)				&$>$1		&MR obs.\\
$\delta$ Cen			&17.9			&$\sim$15 (200)				&$\sim$1&MR obs.\\
$\mu$ Cen					&3.7			&10.4 (144)						&1.03		&-\\
$\alpha$ Ara			&16.0			&15.1 (209)						&1.01		&-\\
\textit{o} aqr		&12.6			&12.8 (177)						&1.00		&Be-Shell\\
\hline
\end{tabular}
\end{table}

\item\underline{$\mu$ Cen} : We obtained two measurements in HR. Because this target is quite faint, i.e., m$_K$=4, the S/N is low. However, ``S'' shaped variations are still clearly present in the differential phases. We also detect a drop of visibility for one baseline, whereas it is clearly below the noise level for the other ones. The line profile is double-peaked and symmetric. The fact that the line intensity is quite low, i.e.  1.3 times the continuum, contributes to the weakness of the visibility drop. The object is unresolved in the continuum.

\item\underline{$\alpha$ Ara} : This star was observed eleven times in LR mode in 2007 but the data quality was too low to obtain more than an estimate of the disk extension (see Sect. 5.1). More recently, we have observed $\alpha$ Ara again during one full night and obtained seven measurements in HR mode. All data exhibit a clear signature of a rotating disk, with a major axis roughly perpendicular to the polarization measurement and compatible with Meilland et al. (2007a) results. The profile is double-peaked and symmetric. The object is partly resolved in the continuum for all baselines, but because of the unstable weather conditions during the observing night, the uncertainties on the absolute visibility remains on the order of 20$\%$.

\item\underline{\textit{o} Aqr} :  This is the faintest target of our sample with m$_K$=4.6. We obtained one measurement in HR mode. The S/N is very low, and we were close to the instrumental sensibility limit considering the quite poor weather conditions during the observations (seeing of 1" and a coherence time of less than 3ms). However, an ``S" phase signal is still clearly visible in the data, at least for the longest baseline. Clues of a small drop of visibility in the line are also present. The profile is typical of Be stars seen at high inclination angle, i.e. double-peaked with a shell absorption line at its center. The object is clearly unresolved in the continuum.
\end{itemize}

\begin{figure*}[!t]
\centering   
\includegraphics[width=0.93\textwidth]{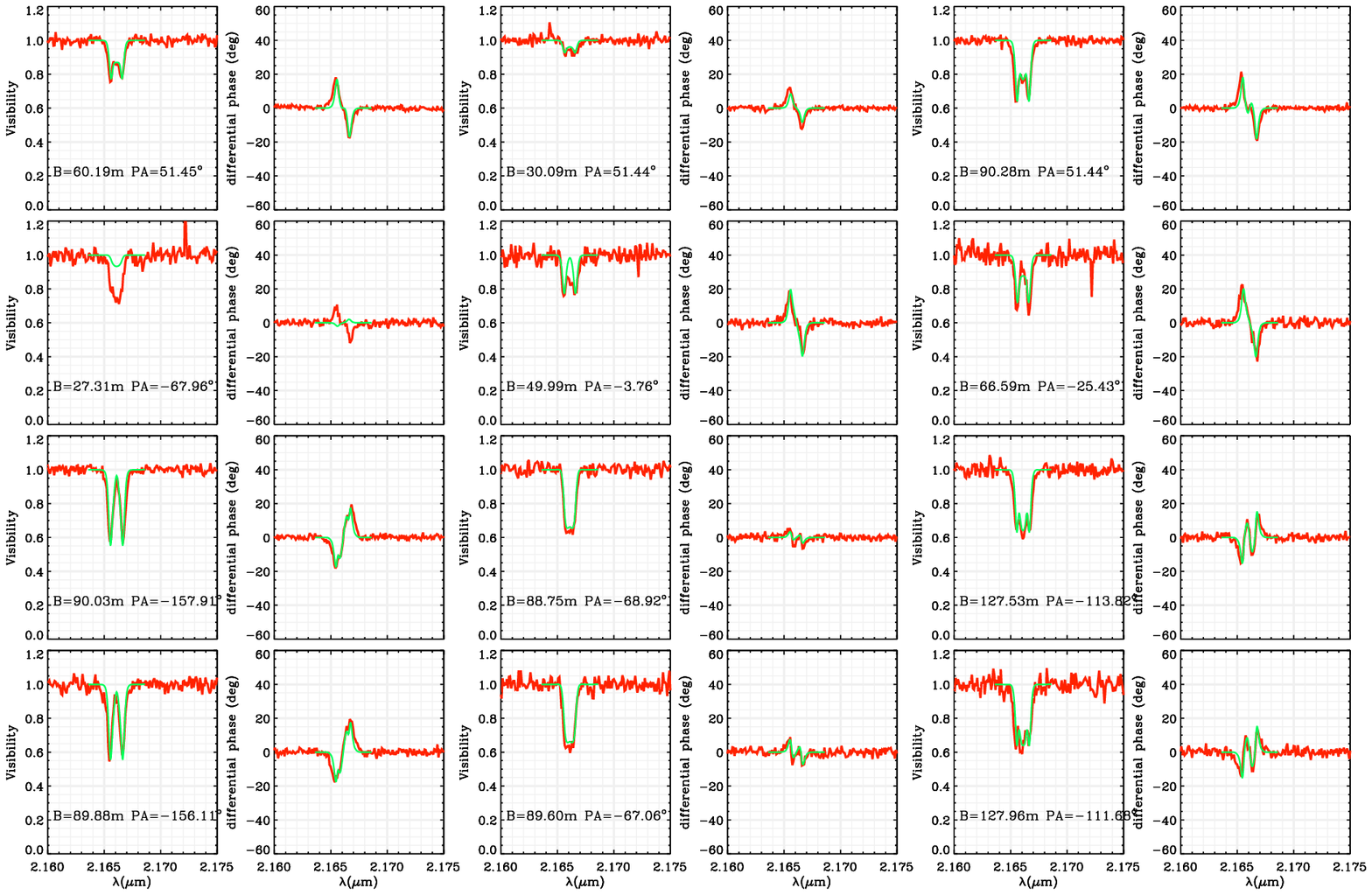}
\caption{$\alpha$ Col differential visibility and phase from our four VLTI/AMBER HR measurements (red line). Each row corresponds to one VLTI/AMBER measurement (three different baselines). The visibility and phase of the best-fit kinematics model is overplotted in green.}
\label{alpcol-vis}
\end{figure*}

\begin{figure*}[!th]
\centering   
\includegraphics[width=0.93\textwidth]{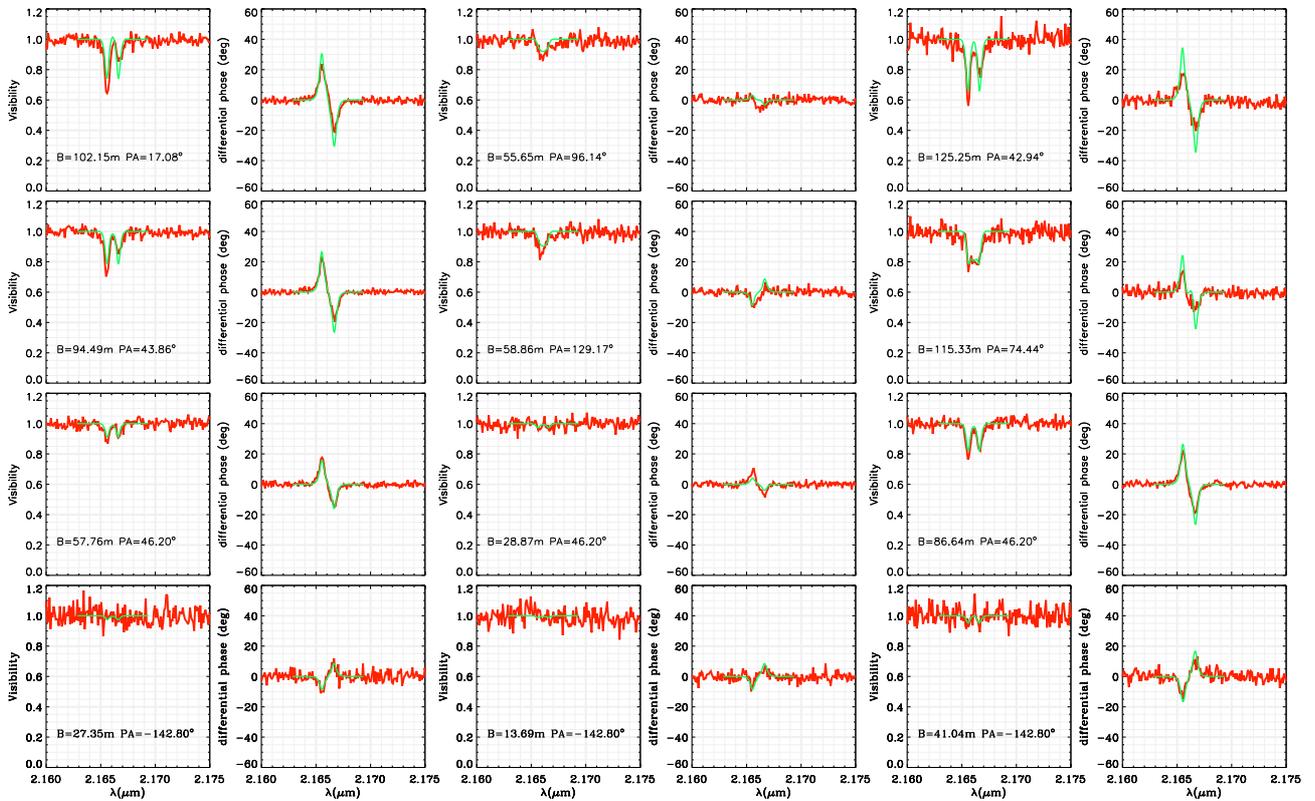}
\caption{$\kappa$ CMa visibility and phase from our four VLTI/AMBER HR measurements (red line). The two first measurements are from 2008 (first two row) and the two other from 2010. The best-fit kinematics model is overplotted in green.}
\label{kapcma-vis}
\end{figure*}

\begin{figure*}[!th]
\centering   
\includegraphics[width=0.89\textwidth]{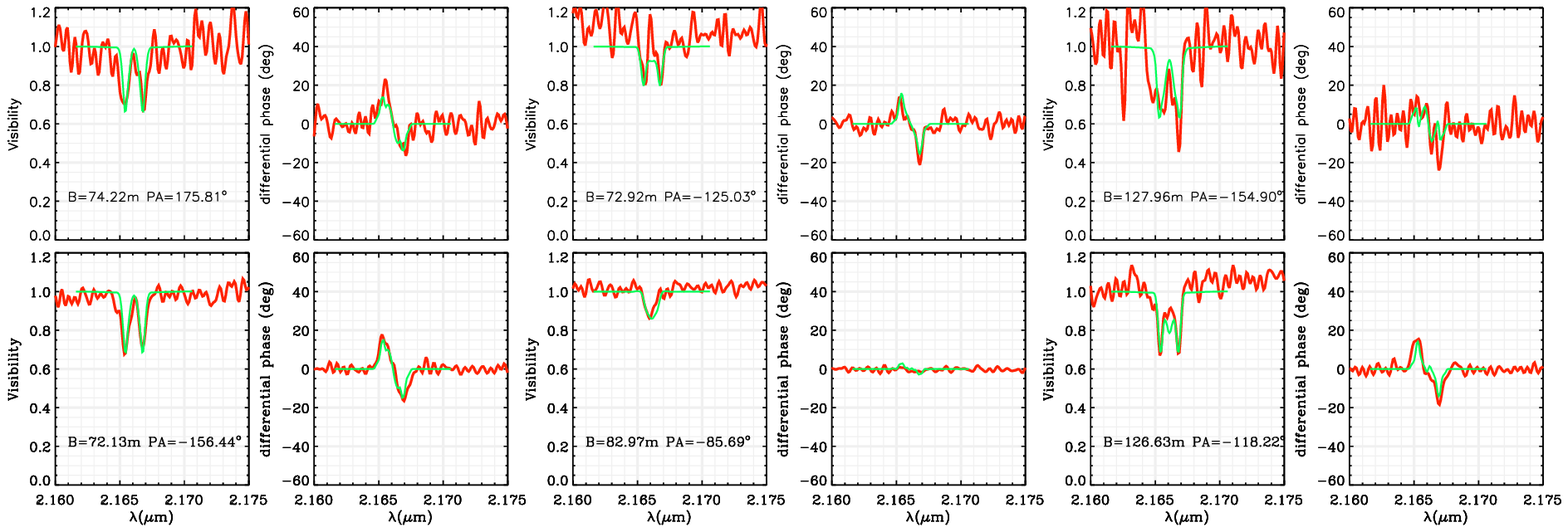}
\caption{$\omega$ Car visibility and phase from our two VLTI/AMBER HR measurements (red line). The  best-fit kinematics model is overplotted in green.}
\label{omecar-vis}
\end{figure*}

\begin{figure*}[!th]
\centering   
\includegraphics[width=0.89\textwidth]{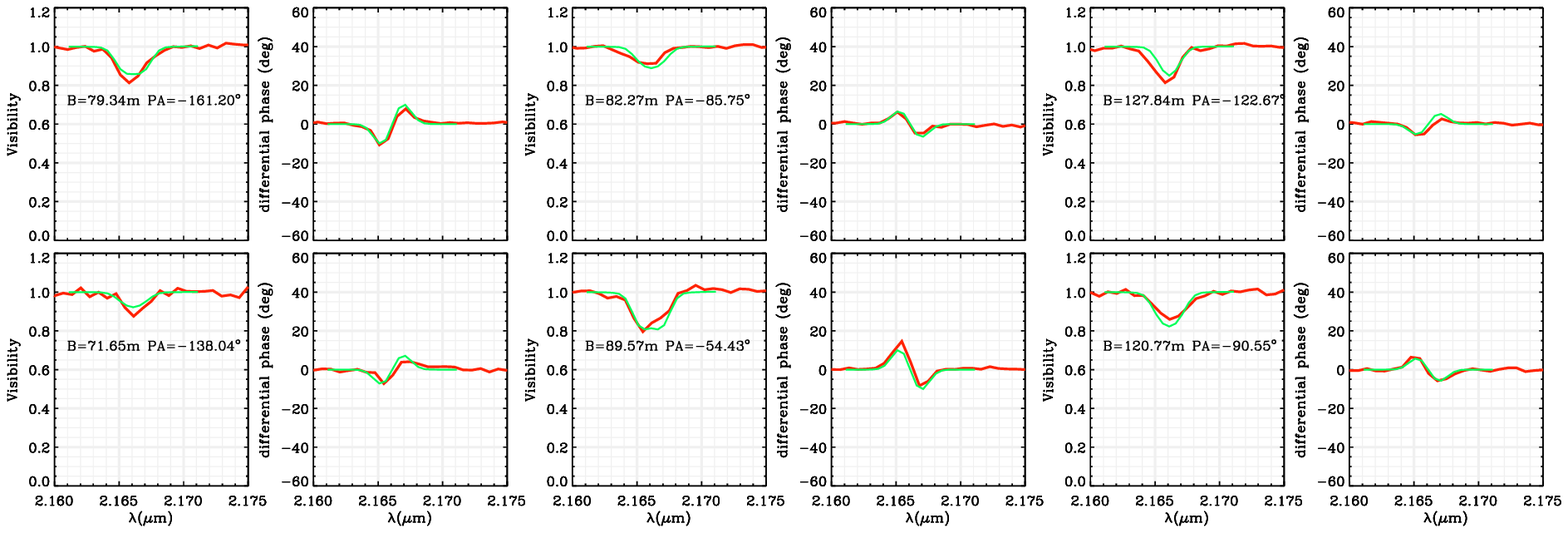}
\caption{p Car visibility and phase from our two VLTI/AMBER MR measurements (red line). The  best-fit kinematics model is overplotted in green.}
\label{pcar-vis}
\end{figure*}

\begin{figure*}[!th]
\centering   
\includegraphics[width=0.89\textwidth]{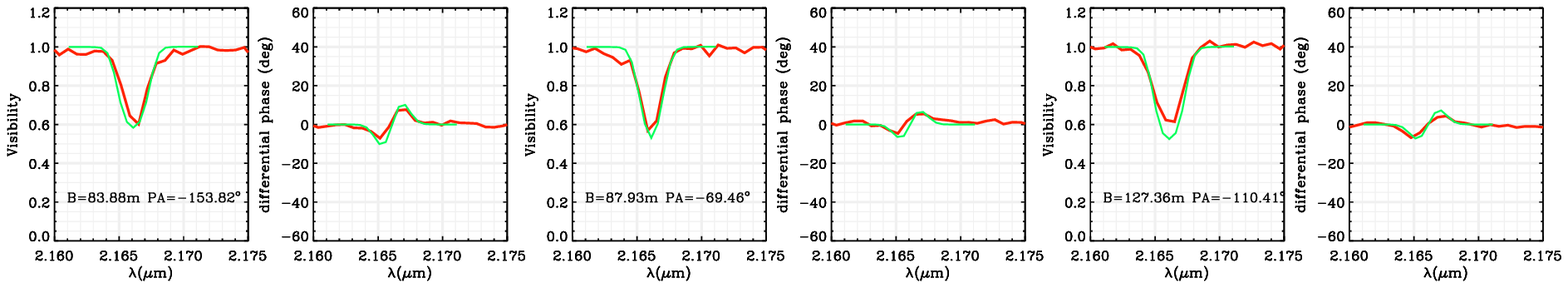}
\caption{$\delta$ Cen visibility and phase from our VLTI/AMBER HR measurement (red line). The  best-fit kinematics model is overplotted in green.}
\label{delcen-vis}
\end{figure*}

\begin{figure*}[!th]
\centering   
\includegraphics[width=0.89\textwidth]{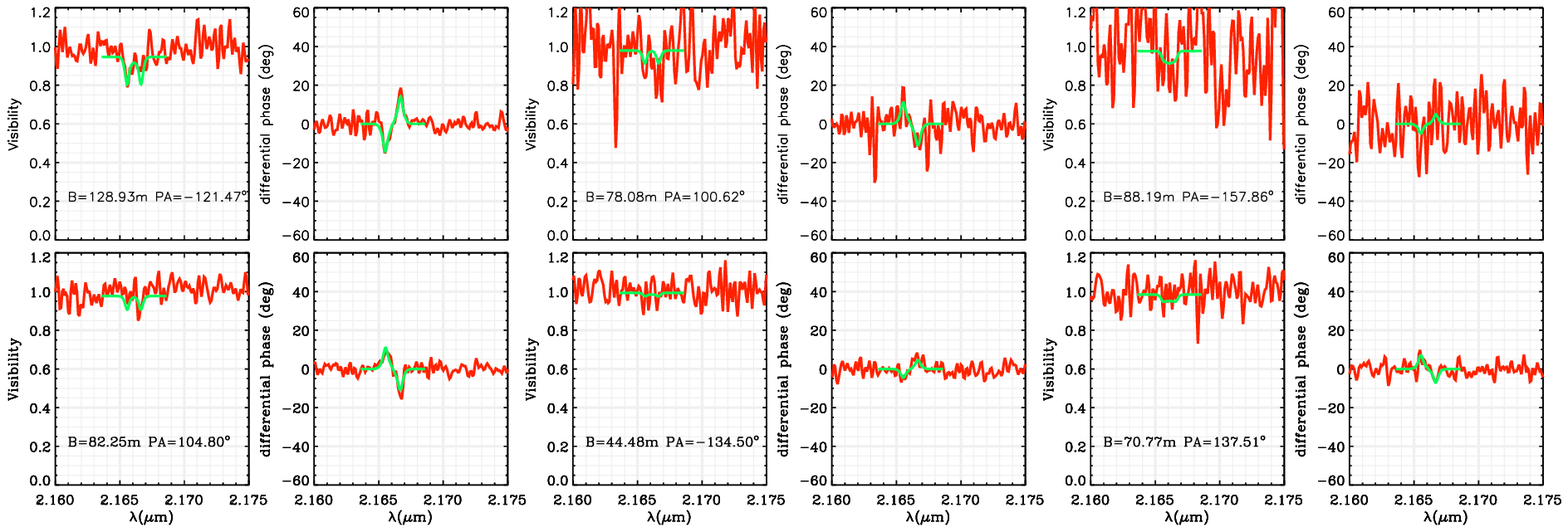}
\caption{$\mu$ Cen visibility and phase from our two VLTI/AMBER HR measurements (red line). The  best-fit kinematics model is overplotted in green.}
\label{mucen-vis}
\end{figure*}

\begin{figure*}[!th]
\centering   
\includegraphics[width=0.88\textwidth]{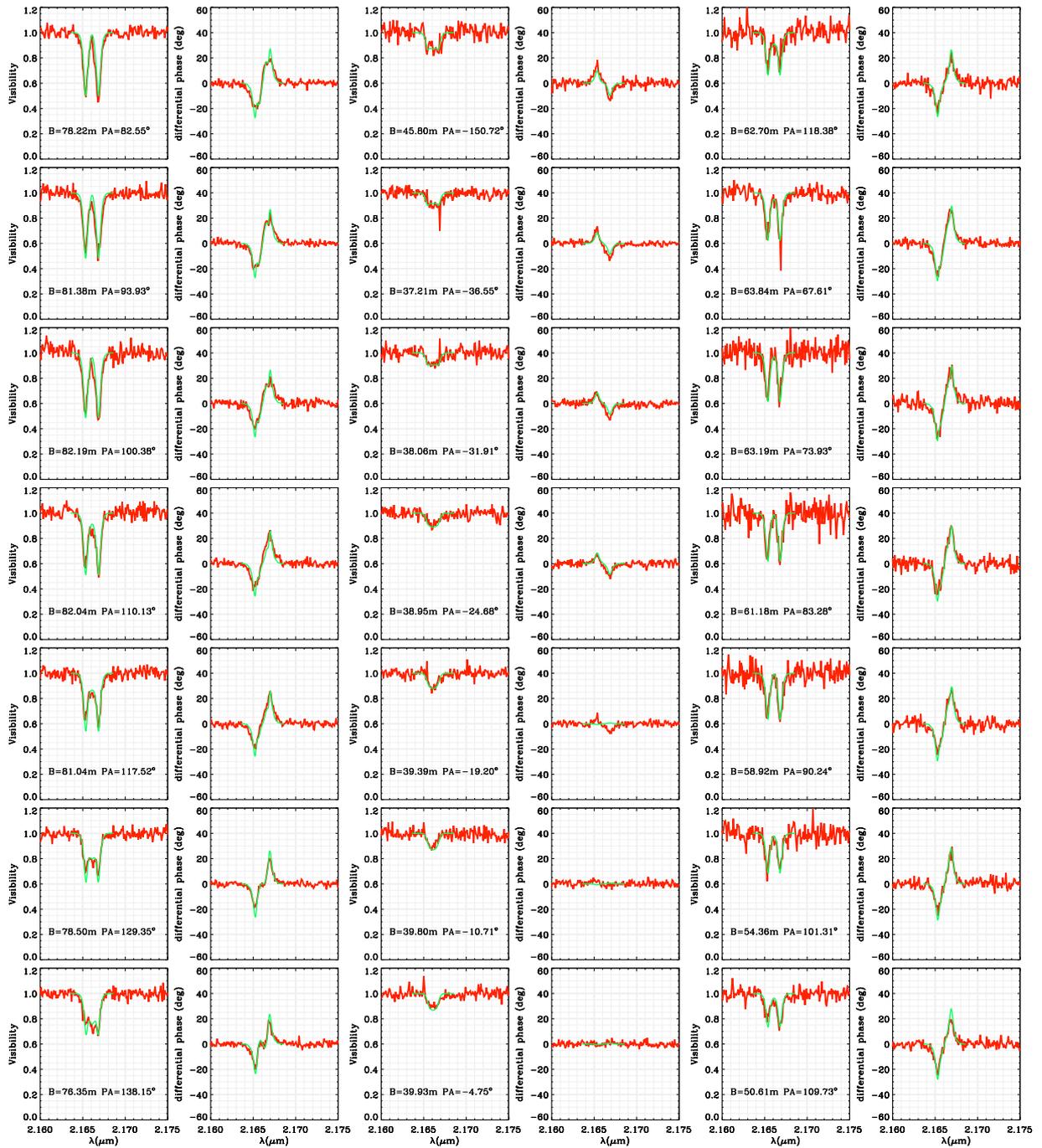}
\caption{$\alpha$ Ara visibility and phase from our seven VLTI/AMBER HR measurements (red line). The best-fit kinematics model is overplotted in green.}
\label{alpara-vis}
\end{figure*}

\begin{figure*}[!th]
\centering   
\includegraphics[width=0.88\textwidth]{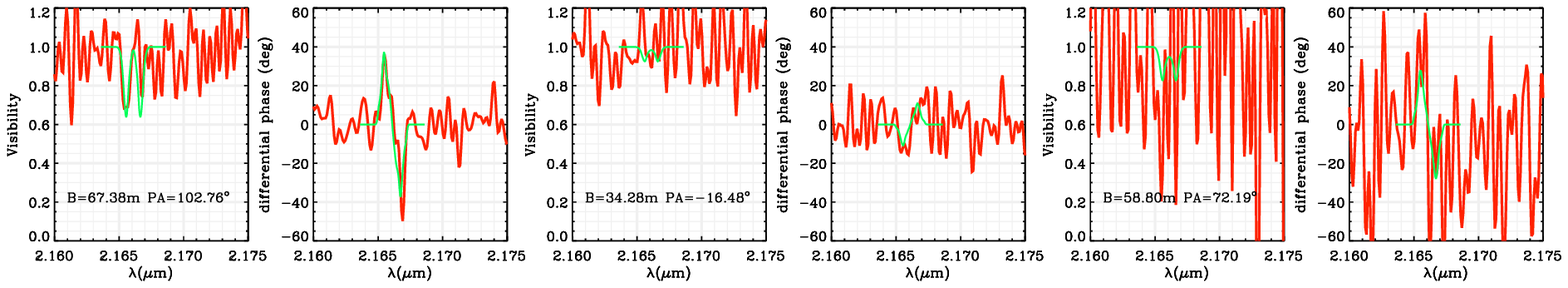}
\caption{\textit{o} Aqr visibility and phase from our VLTI/AMBER HR measurement (red line). The best-fit kinematics model is overplotted in green.}
\vspace{1cm}
\label{omiaqr-vis}
\end{figure*}

\section{Modeling}

In this section we analyze our interferometric data using ``toy models'' of increasing complexity. In the first subsection the K-band continuum visibility is used to estimate the circumstellar environment extension. In the case of the Be binary star $\delta$ Cen we tried to constrain the components' separation at various epochs. Finally, in the last subsection, the differential visibility and phase are used to constrain the circumstellar environment's kinematics.

\subsection{The disk extension in the continuum}

To estimate the circumstellar disk extension in the K-band continuum with our interferometric measurements, we used a simple two components model. Because all  central stars are unresolved or barely resolved even with the longest baselines (i.e. V$>$0.95), they were all modeled as point sources. The second contribution is the circumstellar environment. Owing to the large uncertainties on most of our absolute visibility measurements, we were unable to determine any flattening of the environment, and we decided to simply model this contribution as a circular Gaussian distribution. For the same reason, we decided to set the circumstellar environment relative flux (F$_{\rm env}$) to the value determined from the fit of the SED with a much higher precision than if it had been deduced from the interferometric data only (See Section 2 and Table 1). Finally, the two-dimensional intensity map describing our model is given by

\begin{equation}
I(x,y)=(1-F_{env})\delta(x,y)+\frac{F_{env}}{\sigma\sqrt{2\pi}}exp\Bigl(-\frac{x^2+y^2}{2\sigma^2}\Bigr),
\end{equation}

\noindent where x and y are the Cartesian coordinates, $\delta$(x,y) is the Dirac function, and $\sigma$ is the standard deviation of the Gaussian distribution. In the following, the Gaussian distribution is not defined by its  standard deviation, but by its full width at half maximum defined by FWHM$\simeq$2.35$\sigma$.

For each dataset we separated the low-spectral resolution data from the medium or high-resolution data and fitted the above-described models. The results are summarized in Table~\ref{paramsCont}. In most cases, we see that the accuracy of the LR data is higher than MR or HR, but that stronger biases affect them. Few cases were very problematic and no relevant fit to the data could be obtained. These special cases are marked as "n. c." for "not constrained" in this table.

Using the distance and stellar radius from Table~\ref{params}, we were able to infer the extension in D$_\star$. These values, also presented in Table~\ref{paramsCont}, clearly show that the K-band continuum is quite compact. However, because of the large uncertainties on our measurements, we were unable to determine whether or not the K-band extension depends on any stellar parameters. Nevertheless, we determined a mean size of the environment, i.e. $\overline{\rm FWHM}$=2.2$\,\pm\,$0.3\,D$_\star$ for the whole dataset.

$\alpha$ Ara and $\kappa$ CMa were already observed with the VLTI/AMBER and modeled by Meilland et al. (2007a, 2007b). For $\alpha$ Ara the authors found a mean continuum K-band extension of 6.0$\pm$0.8\,mas, which is significantly higher than our 2.4$\pm$1.1\,mas measurement. However, they used an uniform disk component for the environment. Because, the environment is not fully resolved, we therefore have to apply a x0.87 correcting factor to their measurements to convert the uniform disk diameter into a Gaussian FWHM. Nevertheless, this 6.0x0.87=5.3$\pm$0.7\,mas is still significantly higher than our measurement ($>$2$\sigma$). In the case of $\kappa$ CMa the authors found that the mean K-band continuum extension was lower than 2.7\,mas, which is compatible with our $1.0\pm0.3$\,mas extension.

\begin{table}[!t]
\caption{\label{paramsCont} The K-band continuum extensions of the disks. }
{\centering \begin{tabular}{ccccccc}
\hline 
Name							& F$_{env}$	& \multicolumn{2}{c}{Gaussian FWHM}			&$\chi^2_r$  \\
									& (fixed) 					&  mas										&	D$_\star$		&  \\
\hline\hline  
$\alpha$ Col (HR)	& 0.25							& $1.3\pm0.7$							&1.5$\pm$0.8			&0.7\\
$\alpha$ Col (LR)	& 0.25							& $1.0\pm0.2$							&1.9$\pm$0.4			&2.0\\
$\kappa$ CMa (HR)	& 0.47							& n.c. 										&n.c. 						&n.c.\\
$\kappa$ CMa (LR)	& 0.47							& $1.0\pm0.3$ 						&3.7$\pm$1.1			&4.8\\
$\omega$ Car (HR)	& 0.20							& n.c. 										&n.c. 						&n.c.\\
$\omega$ Car (LR)	& 0.20							& $1.7\pm0.5$							&3.1$\pm$0.9 			&1.9\\
p	Car					(MR)& 0.45							& $1.1\pm0.3$							&2.0$\pm$0.5			&3.1\\
$\mu$ Cen 		(HR)& 0.37							& n.c. 										&n.c.							&n.c.\\
$\alpha$ Ara (HR)	& 0.56							& $2.4\pm1.1$ 						&3.8$\pm$1.7			&7.1\\
$\alpha$ Ara (LR)	& 0.56							& $1.9\pm1.3$ 						&3.0$\pm$2.1			&5.2\\
\textit{o} Aqr(HR)& 0.31							& n.c.										&n.c.							&n.c.\\
\hline 
\end{tabular}\par}
\end{table}

\begin{table}[!b]
\caption{\label{paramsCont deltaCen} Evolution of the $\delta$ Cen component separation.}
{\centering \begin{tabular}{c|cc|cc}
\hline
																																& \multicolumn{2}{c}{Cartesian coord.}	&\multicolumn{2}{c}{Polar coord.}	\\		
Date																														& $\Delta\alpha$		& $\Delta\delta$ 		& sep 			&PA\\
																																&(mas)							&(mas)							&(mas)			&(deg)\\
\hline\hline
2008-01																													&  60.9							& -31.7							&68.7				&117.5\\
2009-03																													& -34.5							& 0.79 							&34.5				&-88.7\\
2010-01																													& -73.8							& 24.5 							&77.8				&-71.6\\
2011-05																													& 2.78							& 9.25 							&9.7				&16.7\\
\hline 
\end{tabular}\par}
\end{table}
\subsection{The binarity of $\delta$ Cen}

We separately fitted the $\delta$ Cen data with a uniform disk + a companion star model to compare it with Meilland et al. 2008. The two-dimensional intensity map is given by 

\begin{equation}
I(x,y)=F_\star\delta(x,y)+F_{comp}\delta(x-\Delta\alpha,y-\Delta\delta)+\frac{4\,F_{env}}{\pi D_{env}^2}\Pi\Bigl(\frac{\sqrt{x^2+y^2}}{D_{env}}\Bigr),
\end{equation}

\noindent where F$_\star$, F$_{comp}$, and F$_{env}$ are the stellar, companion, and environment fluxes, respectively,  ($\Delta\alpha$,$\Delta\delta$) the components separation in Cartesian coordinates, and $\Pi$(t) is the rectangle function defined by $\Pi$(t)=1 for t$\le$1/2 and  $\Pi$(t)=0 for t$>$1/2.  

The fit procedure is similar to that of Millour et al. (2009), i.e. a mix of Levenberg-Marquardt descent with a set of Monte Carlo initial parameters. The only free parameters are the companion coordinates. The other parameters were set to the values determined by Meilland et al. (2008): primary flux F$_\star$=0.41,  envelope flux F$_{\rm env}$=0.52, companion flux F$_{\rm comp}$=0.07, and envelope diameter D$_{\rm env}$=1.6\,mas.  The results are presented in Table~\ref{paramsCont deltaCen}.

We found that the position of the companion varied significantly between the four epochs separated roughly from one year each. It made an almost complete revolution around the main star during this period, pointing to a typical period of the system of about five years. To constrain the orbital elements significantly, the star should be observed again several times with long-baseline interferometry or speckles interferometry. These data should also be completed by radial velocity measurements.

\begin{figure}[!t]
\centering   
\includegraphics[width=0.45\textwidth]{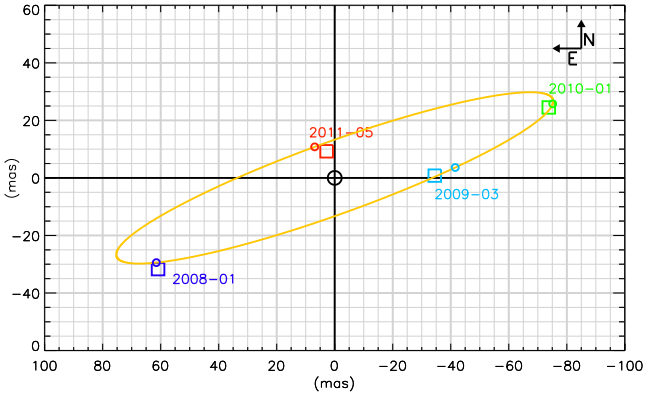}
\caption{$\delta$ Cen binary separation and possible orbit. The measured separations are plotted as squares. The possible orbit is plotted as an orange solid line and the circles represents the modeled positions at the epochs of observations.}
\label{orbit}
\end{figure}

Nevertheless, we tried to determine a first estimate of a possible orbit. It has a probably very low eccentricity, but a very large inclination angle. We managed to obtain a good fit of the orbit with the following parameters: semi-major axis of 80\,mas, periastron in January 2008, period of 5.2\,yr, no eccentricity, inclination angle of 81$^o$, $\omega$= 212$^o$, and $\Omega$=110$^o$. We note that considering the low number of measurements and the probable low eccentricity, this possible orbit, which we overplotted in Fig~\ref{orbit}, may not be unique.

\subsection{The equatorial disk kinematics}

\subsubsection{A simple ``toy'' model}

To quantitatively constrain the velocity fields in the circumstellar environment of the observed Be stars using interferometric observations, we developed a simple two-dimensional kinematic model of a rotating and/or expanding equatorial disk. This model has already been used to model three classical Be stars (see Delaa et al. 2011 and Meilland et al. 2011) and one A[e] supergiant star (Millour et al. 2011) and it is described in detail in Delaa et al. (2011). The model geometry is completely ad-hoc: the star is modeled as a uniform disk and the envelope emission in the continuum and the emission line as two elliptical Gaussian distributions of different FWHMs but with the same flattening due to a projection effect of the geometrically thin equatorial disk, i.e., $f = 1/cos(i)$, where i is the object inclination angle.

\begin{figure*}[!t]
\centering   
\includegraphics[width=0.26\textwidth]{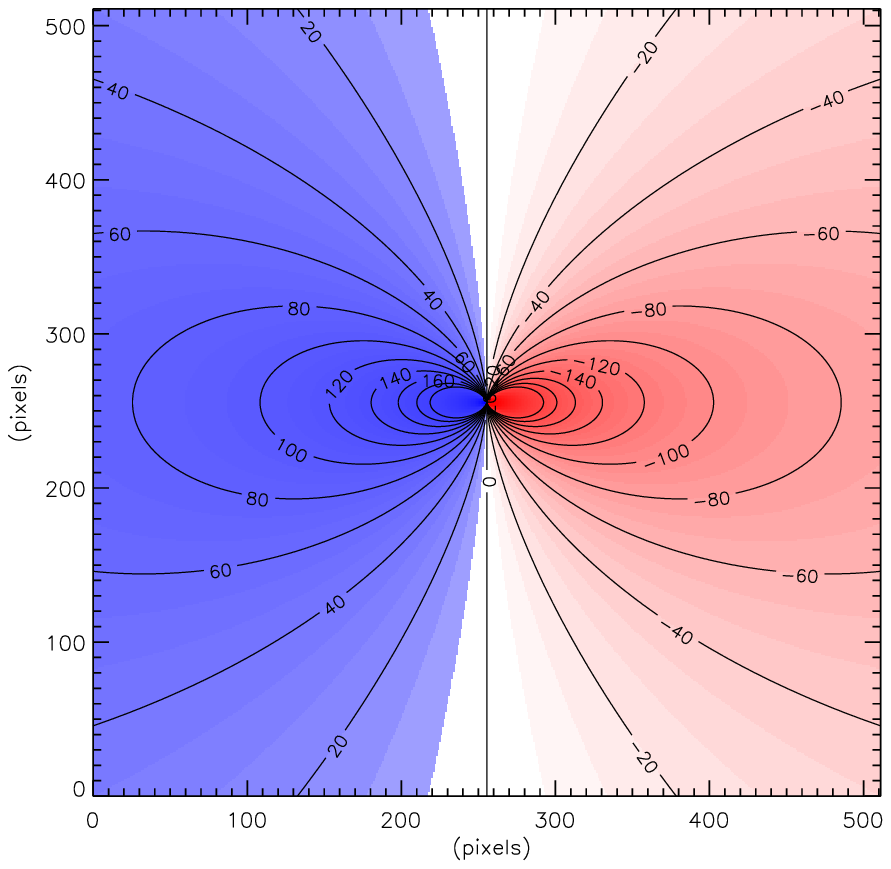}\hspace{1cm}
\includegraphics[width=0.26\textwidth]{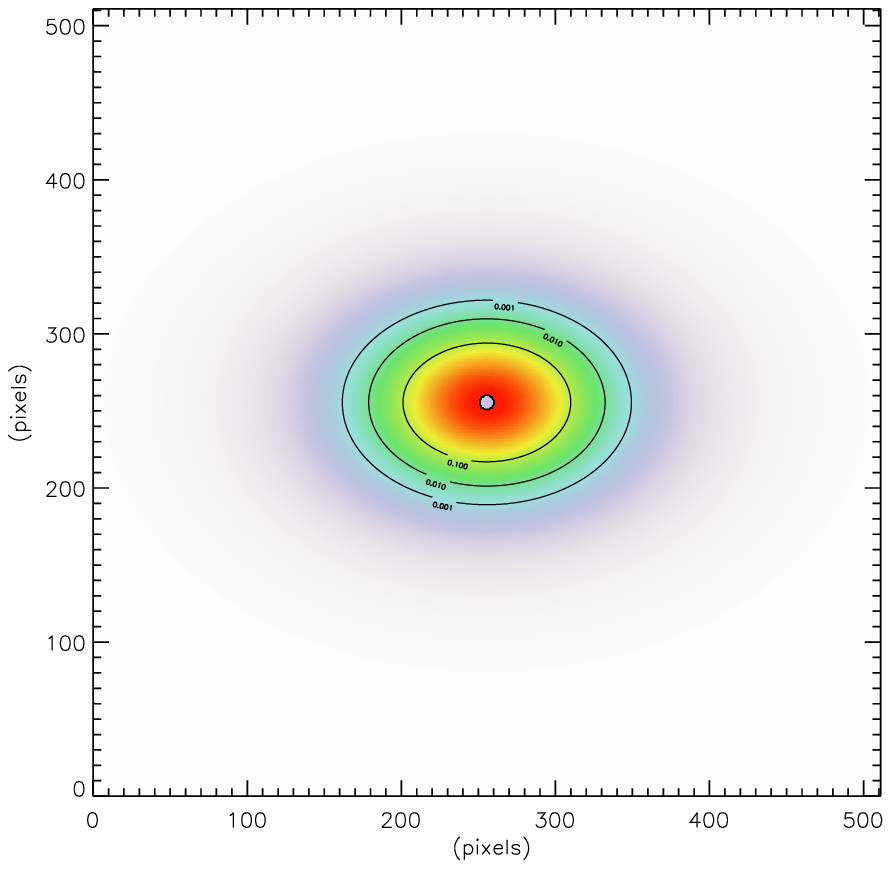}\hspace{1cm}
\includegraphics[width=0.26\textwidth]{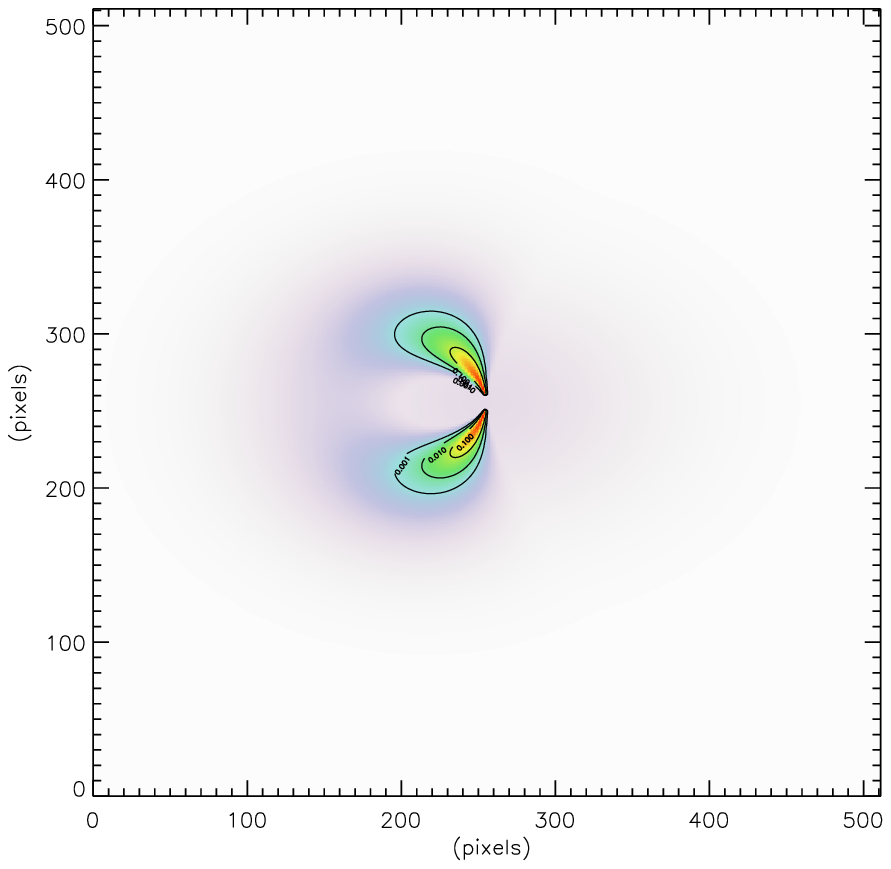}
\vspace{0.3cm}
\caption{Illustration of our kinematic model. Left: two-dimensional project velocity map for a purely rotating disk with an inclination angle of 45$^o$. The blue and red colors represent the positive and negative projected velocities, respectively. Center: global emission map in the Br$\gamma$ line. It is given by a simple elliptical Gaussian with an elongation cause by the projection onto the sky plane. Right: emission map in a narrow spectral channel centered around  21664$\AA$ with a resolution of 1.8$\AA$. It is obtained using the velocity map and the emission map. See Delaa et al. (2011) for more details.}
\label{models}
\end{figure*}

The emission maps were then combined with a two-dimensional projected velocity map of a geometrically thin expanding and/or rotating equatorial disk. For each spectral channel in the line, an iso-velocity map projected along the line of sight was then calculated and multiplied by the whole emission map in the line. Finally, the whole emission map for each wavelength consists of the weighted sum of the stellar map, the disk continuum map and the emission line map within the spectral channel under consideration (see Fig~\ref{models} for an example of emission map obtained in a narrow spectral channel).

The model parameters can be classified into four categories:  

\begin{enumerate}
\item The stellar parameters: stellar radius ($R_\star$), distance ($d$), inclination angle ($i$), and disk major-axis position angle ($PA$).
\item The kinematic parameters: rotational velocity ($V_{\rm rot}$) at the disk inner radius (i.e., photosphere), expansion velocity at the photosphere ($V_0$), terminal velocity ($V_\infty$), and exponents of the expansion ($\gamma$) and rotation ($\beta$) velocity laws. 
\item The disk continuum parameters: disk FWHM in the continuum ($a_{\rm c}$), disk continuum flux normalized by the total continuum flux ($F_{\rm c}$). 
\item The disk emission line parameters: disk FWHM in the line ($a_{\rm l}$) and line equivalent width (EW). 
\end{enumerate}
The star distance is taken from van Leeuwen (2007) and $F_{\rm c}$ and $R_\star$ are derived from the fit of the SED (see Table~\ref{params}). The other nine parameters are free. 

If the disk is directly connected to the stellar surface, the rotational velocity (V$_{\rm rot}$) should be equal to the stellar rotational velocity. However, in some cases, V$_{\rm rot}$ may exceed the stellar velocity if the star is not critically rotating and some additional momentum is transferred to the circumstellar matter. Finally, we considered in our modeling that V$_{\rm rot}$ is free with a higher maximum value equal to the critical velocity (V$_{\rm c}$).

For each target we computed several hundreds of models to constrain the parameters, determined the uncertainties and tried to detect any degeneracy or linked parameters. Owing to the large number of free-parameters, an automatic model-fitting method would have resulted in the computation of millions of models. Moreover, we clearly know each parameter effect on the visibility and phase variations (see Section 5.3.2.). Consequently, we decided to perform the fit manually. For all targets we could exclude models with significant expansion velocity of more than a few km\,s$^{-1}$. Consequently, we decided to set the expansion velocities to zero in all our models. We then tried to constrain the seven remaining parameters ($i$, $PA$, $V_{\rm rot}$, $\beta$, $a_{\rm c}$, $a_{\rm l}$, and $EW_{\rm l}$). To reduce the number of computed models, we started with a qualitative estimation of the parameters from our interferometric data (especially for $PA$, $i$, a$_c$, a$_l$ and $EW_l$) and explored the parameter space with decreasing steps to converge to the $\chi^2$ minimum. To check for other minima, we also explored the full range of possible parameter space but with larger steps. Finally, the parameter values for the best-fit models are presented in Table~\ref{model_params}. The corresponding differential visibilities and phases are overplotted in Figs~\ref{alpcol-vis} to~\ref{omiaqr-vis}. 

The fit quality is very good for three targets observed in HR mode: $\omega$ Car, $\mu$ Cen, and $\alpha$ Ara, and good for the two stars observed in MR mode, i.e., p Car and $\delta$ Cen. It is still satisfying for $\alpha$ Col (i.e. $\chi^2_r$=4), although the visibility and phase of one of the baselines could not be fitted simultaneously with the other ones. In the case of $\kappa$ CMa, the fit is significantly worse  (i.e. $\chi^2_r$=6.8). This is mainly because of the strong asymmetry of this object, which is not taken into account in our simple model. Finally, the data obtained on \textit{o} Aqr seem to be insufficient to fully constrain the model for this object (i.e. $\chi^2_r<$1).

\subsubsection{About the model parameters}

To check the consistency of our modeling, we tried to determine the effects of all model parameters on the visibility and phase variations through the emission line. Some of them strongly affect the interferometric observables and are consequently easily and unambiguously constrained with only a few measurements whereas others are more difficult to infer:

\begin{itemize}

\item the major-axis position angle (\textit{PA}) has a huge effect on the phase variation amplitude and the shape of the visibility drops, as already explained in Meilland et al. (2011). For a non-fully resolved disk, the amplitude of the ``S'' phase variation is proportional to the baseline length, but it also strongly depends on its orientation. The amplitude is maximum for baselines aligned with the major axis and null for the one aligned with the minor axis. For baselines that overresolve the disk, the differential phase loses its simple ``S'' shape and secondary effects become visible (for example see the case of $\alpha$ Col in Fig. 4). The shape of the visibility variations also varies from ``W'' for baselines aligned with the major axis to ``V'' for baselines aligned with the minor axis.

\item the line equivalent width (\textit{EW}$_{\rm l}$) is mainly set by the fit of the line profile. It needs to be corrected by taking into account the photospheric absorption line (see Delaa et al. 2011 for more details).

\item the disk FWHM in the emission line (a$_{\rm l}$) influences the drop of visibility amplitude along all baselines and the amplitude of the ``S'' shape variations. It can also be constrained knowing that the phase variations lose this ``S'' shape for baselines that fully resolved the disk. This parameter also influences the double-peak separation: the larger the disk, the smaller the separation.

\item the disk FWHM in the continuum (a$_{\rm c}$) is mainly derived from the measurements in the continuum presented in Table~\ref{paramsCont}. However, it also indirectly influences the amplitude of the phase variations because it can modify the ratio between the coherent flux (for an unresolved object) and the incoherent one (for a resolved object). The phase variation is proportional to the photocenter shift only for an unresolved object. Therefore, the more resolved an object is in the continuum the smaller will be the phase variations. For example, in the case of $\alpha$ Ara, we did not manage to fit the differential phase with a disk in the continuum that extends to FWHM=3.0\,mas as measured by the absolute visibility (see Table~\ref{paramsCont}), but only with disks smaller than 2\,mas. This may be due to a truncation of the disk, as explained in Chesneau et al. (2005) and Meilland et al (2007a), making it depart from a simple Gaussian shape.

\item the inclination angle (\textit{i}) has a significant influence on the visibility drop amplitude for baselines close to the polar orientation. It is well constrained by comparing equatorial and polar baseline measurements. It also has an effect on the double-peak separations since it influences the projected rotational velocity.

\item the rotational velocity (V$_{\rm rot}$) mainly influences the double-peak separation and the line ``width''. The faster the disk rotates, the larger is the double-peak separation.

\item the exponent of the rotation law ($\beta$) also influences the double-peak separation. With a higher value of $\beta$, the velocity as a function of the distance drops quicker, and the double-peak separation is smaller. Therefore, it is quite hard to distinguish the effect of V$_{\rm rot}$ and $\beta$. However, $\beta$ also influences the shape of wings of the line. The higher the value of $\beta$, the larger the line wings. But, as already mentioned in Delaa et al. (2011), the wings of Be star emission lines can be highly affected by non-kinematic broadening cause by non-coherent scattering, so that it remains hard to set both V$_{\rm rot}$ and $\beta$ unambiguously. Nevertheless, value of $\beta$ of less than 0.3 yields variations that are too sharp, which are not realistic.

\end{itemize}

\begin{table*}[!t]
\caption{Best-fit parameters obtained from our axisymmetric kinematic model.\label{model_params}}
\centering \begin{tabular}{cccccccccc}
\hline
Parameter& unit													&$\alpha$ Col	& $\kappa$ CMa	&$\omega$ Car		&p Car				&$\delta$ Cen	&$\mu$ Cen		&	$\alpha$ Ara	&\textit{o} Aqr\\
\hline\hline
\multicolumn{10}{c}{\textbf{Global geometric parameters}}\\
\hline
$R_\star$&	(R$_\odot)$									&	5.8					&	5.9						&	6.8						&	6.0					&	6.5					&	5.5					&	5.5						&	4\\
$d$&(pc)																&	80					&	202						&	104						&	148					&	127					&	155					&	81						&	133\\
$i$&(deg)  															&	35$\pm$5		&	35$\pm$10			&	65$\pm$10			&	70$\pm$10		&	35$\pm$15		&	25$\pm$5		&	45$\pm$5			&	70$\pm$20\\ 		
$PA$&(deg)															&	10					&	25$\pm$10			&	5$\pm$5				&	-25$\pm$10	&	40$\pm$10		&	80$\pm$15		&	88$\pm$2			&	120$\pm$20\\
\hline
\multicolumn{10}{c}{\textbf{Global kinematic parameters}}\\
\hline
$V_\mathrm{rot}$&(km\,s$^{-1}$) 				&	350$\pm$10	&	480$\pm$40		&	300$\pm$20		&	400$\pm$30	&	500$\pm$50	&	510$\pm$20	&	480$\pm$20		&	400$\pm$50\\
$\beta$&-																&	0.5$\pm$0.1	&	0.5$\pm$0.2		&	0.45$\pm$0.1	&	0.45$\pm$0.1&	0.5$\pm$0.3	&	0.5$\pm$0.1	&	0.5$\pm$0.1		&0.5$\pm$0.2\\
\hline
\multicolumn{10}{c}{\textbf{K-band continuum disk geometry}}\\
\hline
F$_{\rm c}$&-														&	0.25				&	0.5						&	0.2						&	0.45					&	0.45				&	0.37			&	0.56					&	0.31\\
$a_\mathrm{c}$&(D$_\star$)							&	2.$\pm$0.5	&	3.5$\pm$0.5		&3$\pm$1				&	2$\pm$0.5			&	2$\pm$1			&	$<$3			&	$<$2					&	$<$10\\
\hline
\multicolumn{10}{c}{\textbf{Br$\gamma$ disk geometry}}\\
\hline
$a_\mathrm{l}$&(D$_\star$)							&	5.5$\pm$0.3	& 6.5$\pm$2		&	6.5$\pm$1			&	11$\pm$2		&	9$\pm$2			&	4$\pm$1			&	5.8$\pm$0.5		&	14$\pm$1\\
$EW_\mathrm{l}$&($\AA$)								  &	7.0$\pm$0.5	&	13$\pm$2		&	5.8$\pm$0.5		&	10$\pm$1		&	19$\pm$2		&	5.6$\pm$0.3	&	14.5$\pm$1		&	12$\pm$3\\
\hline\hline
\hline
$\chi^2_r$&															& 4.0					&6.8					&1.1						&2.5					&2.3					&1.3					&1.7						&0.8\\
\hline
\end{tabular}
\end{table*}

\section{Discussions}

\subsection{The rotational rate of Be stars}

Using the inclination angle determined from our kinematic model and the v\,sin\,i and the estimate of the critical velocity V$_{\rm c}$ from Fr\'emat et al. (2005), we were able to determine the rotational velocity of the observed targets and constrain the ratio V/V$_{\rm c}$. These results are presented in Table~\ref{vrotable}. We found a mean ratio of $\overline{\rm V/V_c}$=0.82$\pm$0.08. In Fr\'emat et al. (2005), the rotational rate is defined by the ratio of the stellar angular velocity to its critical one :

\begin{equation}
\frac{\Omega}{\Omega_c}=\frac{V}{V_c}\frac{R_{eq\,c}}{R_{eq}}
\end{equation}

\noindent where $\rm{R_{eq}}$ and $\rm{R_{eq\,c}}$ are the equatorial radii (in polar radii) for stars rotating at V and V$_c$, respectively. Under the assumption of the Roche model, $\rm{R_{eq\,c}}$=1.5 and $\rm{R_{eq}}$=1.29 for V/V$_{\rm c}\,\simeq$\,0.82. Consequently, the mean rotational rate for our sample of Be stars is $\overline{\rm \Omega/\Omega_c}$=0.95$\pm$0.02. This value is significantly higher than the one determined by Fr\'emat et al. (2005), i.e. 0.88, from their fit of photospheric lines of Be stars taking into account gravity darkening effects as proposed by Townsend et al. (2004). However, our sample is much smaller than theirs and we note that the inclination angle determined from our modeling agrees with the one determined by Fr\'emat et al. (2005) within 1$\sigma$ for all targets except $\alpha$ Ara (1.3$\sigma$) and $\alpha$ Col (2$\sigma$).

\begin{figure}[!t]
\centering   
\includegraphics[width=0.48\textwidth]{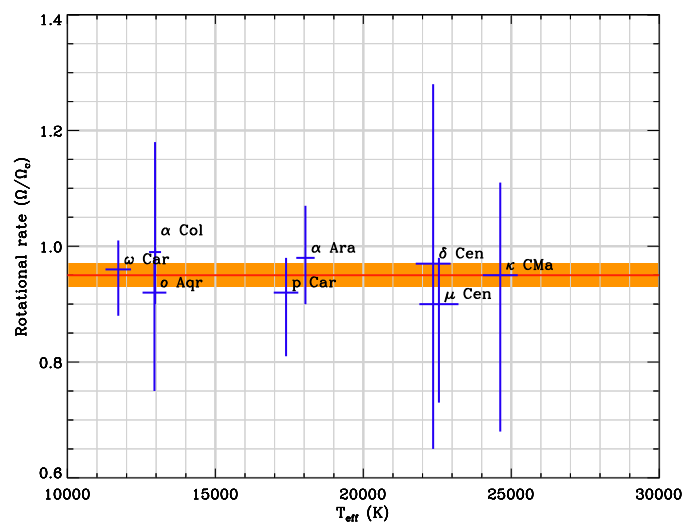}
\caption{Rotational rate of our Be stars sample plotted as a function of their effective temperature. Our measurements for the eight targets and their relative uncertainties are plotted as blue vertical lines. The orange box represents the mean rotational rate, i.e. 0.95$\pm$0.02.}
\label{vrot}
\end{figure}

The values of ${\rm \Omega/\Omega_c}$ for each target are also presented in Table~\ref{vrotable}. We note that the upper value of the uncertainties for some stars are not physically reasonable since stars should not rotate above their critical velocity under the Roche model. However, we assume that the equatorial radius remains equal to 1.5 for stars rotating above V$_{\rm c}$. In Fig.~\ref{vrot}, the Be stars rotational rates are plotted as a function of the effective temperature to see if we find any correlation as proposed by Cranmer (2005). We did not detect any correlation between ${\rm \Omega/\Omega_c}$ and T$_{\rm eff}$. This striking effect could be real, or due to a small sample and our large uncertainties.  

Two stars in our samples have already been studied using the VLTI/AMBER in MR mode: $\alpha$~Ara (Meilland et al. 2007a) and $\kappa$~CMa (Meilland et al. 2007b). The inclination angle found for $\alpha$~Ara, i.e. 45$\pm$5$^o$, is roughly compatible with the previous estimate, i.e. 55$^o$ whereas, for $\kappa$~CMa the two estimates are clearly not compatible, i.e. 35$\pm$5$^o$ in this work and 60$\pm$10$^o$ in Meilland et al. (2007b). The main differences between these previous studies and our current work is that they were conducted at the very beginning of the VLTI/AMBER instrument. At this time, the uncertainties on the absolute visibility measurements were poorly known and had probably been underestimated by the authors. Unlike in the present work, these authors mainly used the absolute visibility measurements to determine the disk flattening and thus infer the inclination angle. Consequently, their conclusion that $\kappa$~CMa rotates at about half its critical velocity is probably biased. Nevertheless, with its strong asymmetry cause by the inhomogeneity in the disk it is still hard to accurately determine the $\kappa$~CMa rotational velocity.

\begin{table}[!b]
\caption{Rotational rate of our Be stars\label{vrotable}}
\centering \begin{tabular}{cccccc}
\hline
Star							&V$_{\rm c}$ 			&V\,sin\,i  			&i						& V/V$_{\rm c}$ &  $\Omega$/$\Omega_{\rm c}$ \\
									&km\,s$^{-1}$			&km\,s$^{-1}$			&	deg					&								&	\\
\hline\hline
$\alpha$ Col			&355$\pm$23				&192$\pm$12				&35$\pm$5			&0.95$\pm$0.23	& 0.99$^{+0.19}_{-0.09}$\\
$\kappa$ CMa			&535$\pm$39				&244$\pm$17				&35$\pm$10		&0.80$\pm$0.31	& 0.95$^{+0.16}_{-0.27}$\\
$\omega$ Car			&320$\pm$17				&245$\pm$13				&65$\pm$10		&0.84$\pm$0.16	& 0.96$^{+0.05}_{-0.08}$\\
p Car							&401$\pm$28				&285$\pm$20				&70$\pm$10		&0.76$\pm$0.15	& 0.92$^{+0.06}_{-0.11}$\\
$\delta$ Cen			&527$\pm$29				&263$\pm$14				&35$\pm$15		&0.87$\pm$0.41	& 0.97$^{+0.31}_{-0.32}$\\
$\mu$ Cen					&508$\pm$32				&155$\pm$4				&25$\pm$5			&0.72$\pm$0.20	& 0.90$^{+0.08}_{-0.17}$\\
$\alpha$ Ara			&477$\pm$24				&305$\pm$15				&45$\pm$5			&0.90$\pm$0.17	& 0.98$^{+0.09}_{-0.08}$\\
\textit{o} Aqr		&391$\pm$27				&282$\pm$20				&70$\pm$20		&0.77$\pm$0.21	& 0.93$^{+0.06}_{-0.17}$\\
\hline
\end{tabular}
\end{table}

\subsection{The equatorial disk extension}

All objects we studied are, at least, partly resolved in the Br$\gamma$ line. We managed to significantly constrain the extension of the line emission for all targets. We found Gaussian FWHMs that range between 4 to 14 stellar diameters. We found no correlation between the Br$\gamma$ emission and other properties or characteristics except for the double-peaked separation (see the next subsection for the discussion on the disk kinematics.). It seems that the size is independent of the stellar parameters and of the infrared excess or line equivalent width. The mean FWHM of the Br$\gamma$ line emission is 6.1\,$\pm$\,2.9\,D$_\star$

As already explained in Sect. 5.1, most of the targets are also partly resolved in the continuum with FWHM that range between 1.5 and 3.7\,D$_\star$ and a mean FWHM of 2.2\,$\pm$\,0.3\,D$_\star$. This is roughly compatible with K'-band CHARA interferometer measurements by Gies et al. (2007) on the Be stars $\gamma$ Cas (2.4\,D$_\star$), $\phi$ Per (3.3\,D$_\star$), $\zeta$ Tau (5.5\,D$_\star$), and $\kappa$ Dra (4.3\,D$_\star$). 

These typical K-band continuum and Br$\gamma$ line extensions are significantly smaller than the disk size measured in H$\alpha$. For example a multi-line spectro-interferometric study of the circumstellar environment of the Be star $\delta$ Sco published in Meilland et al. 2011 shows that the emission extension was 1.6 times more extended in H$_\alpha$ (9\,D$_\star$) than in Br$\gamma$ (5.5\,D$_\star$) . Other narrow band studies in H$\alpha$ published in Tycner et al. (2004, 2006) and Quirrenbach et al. (1997) also concluded that the typical H$\alpha$ extension were of the same order.

 As explained in Meilland et al. (2007a, 2007b) using the SIMECA code (Stee et al. 1994) and Tycner et al. (2007) using the BEDISK code (Sigut \& Jones 2007), these interferometric measurements can be used to constrain the circumstellar environment physical parameters:  mass loss, disk mass, and temperature and density distribution. In a forthcoming paper, we will use these numerical codes and the available R (CHARA/VEGA), K (VLTI/AMBER), and N (VLTI/MIDI) band measurements to draw a more complete picture of the Be star circumstellar environments.

\subsection{The equatorial disk kinematics}

For all our targets, the simple kinematic model reproduced our VLTI/AMBER measurements  very well. Therefore, it is clear that most of the Br$\gamma$ line emission comes from the equatorial disk. The disk kinematics is dominated by rotation, with a rotational law close to Keplerian for all targets. The putative expansion velocity is far below the detectability limit of the instrument (i.e. $<$10 km\,s$^{-1}$). These results fully agree with previous kinematics studies performed with the VLTI/AMBER (Meilland et al. 2007a, 2011) or CHARA/VEGA (Delaa et al. 2011). 

If the disks are in Keplerian motion and the stars are rotating significantly below their critical velocity, two questions remain: 
\begin{itemize}
\item What additional mechanisms provide the amount of energy needed to launch the matter from the stellar surface?
\item How does the ejected matter gain sufficient kinetic energy to accelerate up to the Keplerian velocity? 
\end{itemize}

In other terms, the matter needs to be accelerated both radially and azimuthally. Lee et al. (1991) proposed that the disk could be formed by the effect of the gas viscosity that drifts the matter outward. However, in his theory, a source to supply angular momentum at the stellar surface is still needed if the star is not critically rotating. Non-radial pulsations as proposed by Osaki (1986) could be a good candidate. Recently, Cranmer (2009) proposed a theory in which  resonant oscillations in the photosphere could inject enough angular momentum to spin up a Keplerian disk even for the slowest rotating Be stars (down to 60$\%$ of V$_{\rm c}$). However, the question remains open and other mechanisms such as magnetism (Yudin et al. 2010), radiative pressure (Abbott 1979) or binarity could also contribute to the ejection of matter. To solve the question, the connecting layers between the stellar surface and the inner part of the equatorial disk should be carefully studied. This could be done by studying many absorption and emission lines formed close to the photosphere and by comparing their morphology.

\subsection{Departures from our simple model}

Despite the generally good fit of our interferometric data using the simple kinematic model, there are two significant departures from the model that need to be investigated:

\begin{itemize}

\item The most important one is the case of $\kappa$ CMa that is clearly showing an asymmetric profile and visibility and phase variations. These asymmetries can only originate from inhomogeneities in the circumstellar environment. To determine whether or not these inhomogeneities can be modeled as one-armed oscillation as proposed by Okazaki (1997) a dedicated model needs to be developed. Such a thorough analysis of $\kappa$ CMa data is out of the scope of this paper. We note that despite the lower resolution, the p Car data are also showing clues of the same kind of asymmetries.

\item The second most important departure from the model concerns the measurement obtained on $\alpha$ Col with the short polar baseline that cannot be fitted simultaneously with the other ones. This measurement shows that the environment is almost fully resolved with B$\simeq$27\,m for this orientation, whereas it is clearly less extended with a baseline closer to the equator. This could be a clue of a polar wind, as already detected by Kervella \& Domiciano de Souza (2006) on the classical Be star Achernar. However, a detailed study of the circumstellar environment of Be stars along their polar axis is needed to definitively answer the question of the detectability of polar winds in the K-band (see Stee 2011 for a detailed discussion on the polar winds of Be stars).
\end{itemize}

These two questions concerning the circumstellar environments of Be stars will be studied in detail in some forthcoming dedicated papers.

\section{Conclusion}

In this spectro-interferometric survey of classical Be stars we managed to resolve all targets, constrain their extension in the Br$\gamma$ line and, for some of them, in the K-band continuum. Using a simple kinematic model of a purely rotating disk, we were able to successfully model all our data, showing that most of the Br$\gamma$ emission originates from the equatorial region. The disk is fully dominated by rotation, and the rotation law was found to be  Keplerian or quasi-Keplerian for all targets. We were also able to significantly constrain the stellar rotational velocity with our estimates of the inclination angle. We found a mean rotation rate of $\overline{\rm \Omega/\Omega_c}$=0.95$\pm$0.02. With such a high value of the rotational rate, only a small amount of energy is needed to expel matter from the stellar surface. This conclusion is quite different from the estimate by Fr\'emat et al.(2005) obtained by modeling photospheric lines.

We did not detect any correlation between the stellar parameters and the disk properties. However, the uncertainties on the measurements remain high and our sample of Be stars needs to be extended to definitively answer the question on the physical process or processes responsible for the mass ejection and the dependence of the Be phenomenon on the stellar parameters.

In a forthcoming paper, these data will be analyzed using the radiative transfer codes SIMECA and BEDISK to fully constrain the circumstellar environment.

\begin{acknowledgements}
The Programme National de Physique Stellaire (PNPS) and the Institut National en Sciences de l'Univers (INSU) are acknowledged for their financial supports. S. Kanaan also acknowledges financial support from the GEMINI-CONICYT Fund, allocated to the project N° 32090006".
\end{acknowledgements}

\end{document}